\documentstyle[pre,aps,floats,twocolumn,epsf]{revtex}

\begin{document}

\def\d{\mbox{$d$}}
\def\N{\mbox{$N$}}
\def\R{\mbox{$I\!\!R$}}

\def\A{\mbox{\bf A}}
\def\B{\mbox{\bf B}}
\def\C{\mbox{\bf C}}
\def\D{\mbox{\bf D}}
\def\E{\mbox{\bf E}}
\def\F{\mbox{\bf F}}
\def\G{\mbox{\bf G}}
\def\K{\mbox{\bf K}}
\def\M{\mbox{\bf M}}
\def\P{\mbox{\bf P}}
\def\Q{\mbox{\bf Q}}
\def\U{\mbox{\bf U}}
\def\DF{\mbox{\bf DF}}
\def\DE{\mbox{\bf DE}}

\def\e{\mbox{$\hat{{\bf e}}$}}

\def\u{\mbox{\bf u}}
\def\w{\mbox{\bf w}}
\def\x{\mbox{\bf x}}
\def\y{\mbox{\bf y}}
\def\z{\mbox{\bf z}}

\def\Tr{\mbox{\text Tr}}

\def\zero{\mbox{\bf 0}}

\draft

\title{Synchronization of chaotic systems: Transverse stability
of trajectories in invariant manifolds}

\author{Reggie Brown and Nikolai F. Rulkov}
\address{Institute for Nonlinear Science, University of
California, San Diego, La Jolla, CA 92093-0402}

\date{\today}
\maketitle

\begin{abstract}
We examine synchronization of identical chaotic systems coupled
in a drive/response manner.  A rigorous criterion is presented 
which, if satisfied, guarantees that synchronization to the driving
trajectory is linearly stable to perturbations.  An easy to use
approximate criterion for estimating linear stability is also 
presented.  One major advantage of these criteria is that, for  
simple systems, many of the calculations needed to implement them 
can be performed analytically.  Geometrical interpretations of
the criterion are discussed, as well as how they may be used to
investigate synchronization between mutual coupled systems and 
the stability of invariant manifolds within a dynamical system.
Finally, the relationship between our criterion and results from
control theory are discussed.  Analytical and numerical results 
from tests of these criteria on four different dynamical systems
are presented.

\end{abstract}


{\bf If several identical dynamical systems travel the
same phase space trajectory at the same time then
they are said to be synchronized.  Examples of this
type of behavior go back at least as far as Huygens.
Surprisingly, this type of behavior has also been
observed for chaotic systems.  If the systems are 
chaotic then they must be coupled together in some
fashion.  A typical procedure for determining a type
of coupling that produces synchronous motion is to 
try different couplings until one of them works.  
This paper presents a rigorous and an approximate
criteria that, if satisfied, guarantees linearly 
stable synchronous motion for the systems.  To use
the criterion one only needs the equations of motion
and the trajectory one wants the systems to follow.
The paper also presents a step by step procedure 
that will allow one to design couplings that satisfy
these criteria.  Since the procedure does not use
trial and error, and (for simple systems) many of
the steps can be performed analytically, it is an
improvement over current practice.}

\section{Introduction}
\label{intro}
Since Fujisaka and Yamada's 1983 paper on synchronized motion in
coupled chaotic systems many researchers have discussed the
stability of this type of motion~\cite{fy}.  Rather than attempt to 
reference a complete list of papers on this topic we refer the
interested reader to Refs.~\cite{cp,rnn,spie,uf,sushchik,hcp,gb,cd}.
(These papers have extensive bibliographies and represent a
reasonable introduction to the literature.)
The discussion in this paper will center around the type of
synchronization discussed by most of these authors.  Namely, two
or more {\em identical} chaotic systems, coupled in a 
drive/response manner, which exhibit motion that is chaotic and
{\em identical} in time.   (Although the dynamical systems we
examine are chaotic our results are applicable to nonlinear
systems that exhibit other types of motion.) 

The central question addressed in this paper is: ``Given two arbitrary
identical dynamical systems how can one {\em design} a physically
available coupling scheme that is guaranteed to produce stable 
synchronized motion.''  Despite the large amount of effort devoted
to this issue there are relatively few rigorous results.  In most
cases rigorous results are obtained using Lyapunov 
functions~\cite{hv,hmr,nr,wu}.  Unfortunately, this method is not
regular since, in practice, it can only be applied to particular
examples.  I.e., given an arbitrary dynamical system it is not 
clear how one can derive a coupling scheme and/or a Lyapunov function
which guarantees stable synchronous motion.  

Another rigorous approach is that of Ashwin et. al.~\cite{abs}.
In a series of
papers they presented results that are related to the central
question.  In their work the measure (equivalently, the trajectory)
used to describe the dynamics of the system is of central importance.  
To apply the approach one must show that all normal Lyapunov exponents
are negative for all measures of the dynamics.  Unfortunately, it is
well known that Lyapunov exponents can be difficult to calculate.  
Non-rigorous results that employ Lyapunov exponents to determine the 
stability of measures to perturbations have also appeared in the
literature~\cite{hcp,gb,pc,sll,pddy}.

A third rigorous approach by Walker and Mees has recently 
appeared~\cite{wm}.  They uses the method of Lyapunov to develop a
sufficient condition for stable synchronization.  Their condition
relates eigenvalues of the linear part of the vector field to the
Jacobian of the nonlinear part.  The analysis in this paper also d
differs from that in other papers in that they interpret 
synchronization as an observer problem.  

Finally, there are a few special cases where, due to the coupling 
between the nonlinear systems, rigorous analysis of the stability
of synchronization is straightforward.  One case is when the coupling 
transforms the driven system into a stable homogeneous linear system
with time dependent external forcing, (see Ref.~\cite{rvrrv94}).
A second is when the coupling between all of the variables is 
diagonal~\cite{fy,hcp}.  In many practical cases these types of 
coupling can't be achieved.

The approach advocated in this paper is like that of
Ashwin~et.~al. in that it emphasizes the role of individual
trajectories within the dynamics of the system.  However, we
do not explicitly use Lyapunov exponents or Lyapunov functions to
examine the linear stability of synchronous motion.  It is also
like the paper by Walker and Mees in that their stability condition
and design concepts are similar the one we develop below.  The
issue of design has also been addressed by Peng~et.~al.~\cite{pddy}.

We have two major results.  The  first is a rigorously
derived criterion which, if satisfied, {\em guarantees}  that the
coupling scheme will yield linearly stable synchronous motion 
on the driving trajectory.  The second result is a
simple, ``quick and dirty'' criterion that can be used to
estimate the coupling strength needed for linear stability.
This second criterion is
easy to implement and can be used to quickly examine a large
range of coupling schemes and strengths.  For both cases,
linear stability is with respect to perturbations that are
transverse to the synchronization manifold. (Of course, because 
our results are obtained from a linearization they can not fully
address the complications that arise when nonlinear effects are
incorporated.  Some of these difficulties are
discussed in Section~\ref{sum}.)

References~\cite{gb}, \cite{abs}, \cite{hcp2}, \cite{os} and 
\cite{ga}, and the results presented
below, indicate that the stability of the synchronization manifold
depends on the measure used to describe the dynamics on the
manifold.  If the synchronization manifold contains a set of
chaotic trajectories then there are an infinite number of
possible measures for the dynamics on the manifold.  Under these
circumstances a practical answer to the question of the stability
of the synchronization manifold can be quite complicated.  For 
example, the manifold may be linearly unstable when one uses a 
measure confined to one periodic orbit on the manifold (a Dirac
measure~\cite{abs}) and linearly stable on a measure confined to
a different periodic orbit.  At the end of the next section we
discuss relevant results that may allow one to overcome this 
difficulty.

We close this section with an outline of the remainder of this
paper.  In Sections~\ref{theory1} and \ref{theory2} we derive 
criteria that indicate when synchronization between
drive and response systems will occur.  The results in
Section~\ref{theory1} are rigorous while those in
Section~\ref{theory2} are approximations.  Section~\ref{results}
presents results from analytic examinations and numerical
experiments we have conducted on the Rossler, Lorenz, and  
Ott--Sommerer systems.  In Section~\ref{sum} we summarize our 
results and discuss how nonlinear effects impact our results. 
In Appendix~\ref{other} we discus how our results apply to other 
systems (in this appendix we examine a system of equations
that come about when considering chaotic masking for private 
communications).  In Appendix~\ref{geometry1} and we present 
geometric interpretations of the manifolds that arise in our
discussion, while in Appendix~\ref{geometry2}
we discuss a geometrical interpretation of the theoretical
results.  In Appendix~\ref{control} we discuss the relationship
between our results and control theory, while Appendix~\ref{lorenz}
is devoted to an alternative analysis of our criterion and an
additional examination of the Lorenz system.

\section{Theory Part I:  Rigorous Results}
\label{theory1}
This section presents rigorous results regarding linear stability
of trajectories on the invariant manifold associated with
synchronous motion.  (Here, and for the remainder of this paper,
when we say synchronization to a trajectory is linearly stable we 
mean it is linearly stable with respect to perturbations that are 
transverse to the part of the synchronization manifold defined by 
the driving trajectory.)  The major result is a criterion which, 
if satisfied, guarantees linear stability of the driving trajectory
within the synchronization manifold.  The criterion can be used to
design couplings that guarantee linearly stable synchronization.  

We will explicitly examine one type of
drive/response coupling for identical systems that
have chaotic uncoupled dynamics~\cite{od,nagk,RVSPIE}.
Assume the dynamics of the driving system is given by
\begin{equation}
\label{drive}
\frac{d \x}{dt}  =  \F(\x; t)
\end{equation}
where $\x \in \R^d$.  In the presence of coupling the dynamics of
the response system becomes
\begin{equation}
\label{response}
\frac{d \y}{dt} = \F(\y;t) + \E (\x - \y),
\end{equation}
where \E\ is a vector function of its argument and represents
coupling between the systems.  We assume $\E(\zero) = \zero$, hence
synchronization occurs on the invariant
manifold given by $\x =  \y$.  Obviously,
if the coupling strength is below some critical threshold then
synchronization will not occur.  In addition, for some \F's and
choices of \E, synchronous
motion only occurs within some finite range of coupling
strengths.  For these situations if the coupling strength is too
small {\em or} too large then synchronization will not occur. 
Finally, for some \F's and choices of \E\ synchronization will never
occur.

Since we are interested in deviations of \y\ from \x\ we use
Eqs.~(\ref{drive}) and (\ref{response}) to obtain the following
linearized equation of motion for $\w \equiv \y -\x$ (motion
transverse to the synchronization manifold)
\begin{equation}
\label{linear}
\frac{d \w}{dt} = \left[ \DF(\x ;t) - \DE(\zero) \right] \w.
\end{equation}
In this equation $\DF(\x ;t)$ is the Jacobian of \F\ evaluated at \x\
at time $t$, and $\DE(\zero)$ is the Jacobian of \E\ evaluated at $\w =
\zero$.  The synchronization manifold is linearly stable if 
\begin{displaymath}
\lim_{t \rightarrow \infty} \| \w (t) \| = \lim_{t \rightarrow \infty}
\| \y(t) - \x(t) \| = 0
\end{displaymath}
for all possible driving trajectories, $\x(t)$.  (Notice that a
stability analysis of synchronized chaos needs to consider only the
invariant trajectories, $\x(t)$, that belong to the chaotic attractor
of the driving system.)  We begin determining the behavior of $\w(t)$ in 
this limit by dividing 
$\DF(\x; t) - \DE(\zero)$ into a time independent part, \A, and an
explicitly time dependent part, $\B(\x ;t)$,
\begin{equation}
\label{decomp}
\DF(\x; t) - \DE(\zero) \equiv \A + \B(\x; t).
\end{equation}

This decomposition is not unique since we are free to
add constant terms to \A\ provided we are willing to
subtract the same term from \B.  This freedom will be
resolve later when an explicit decomposition is determined.  For
now the reader is asked to accept Eq.~(\ref{decomp}) as a formal
decomposition.
Denote, and order, the eigenvalues of \A\ by $\Re[\Lambda_1] \geq
\Re[\Lambda_2] \geq \cdots \geq \Re[\Lambda_d]$, where
$\Re[\Lambda]$ is the real part of the eigenvalue $\Lambda$.  
Associated with the eigenvalues are eigenvectors denoted by
$\e_1, \,  \e_2, \, \cdots \e_d$.   Next, assume \A\ can be
diagonalized by $\P = \left[ \e_1 \, \e_2 \, \cdots \e_d
\right]$.  Thus, $\D = \P^{-1} \A \P$ is a diagonal matrix.
Rewriting Eq.~(\ref{decomp}) in terms of the coordinate system 
defined by the eigenvectors of \A\ yields
\begin{displaymath}
\frac{d \z}{dt} = \left[ \D + \K(\x; t) \right] \z,
\end{displaymath}
where $\z = \P^{-1} \w$ and $\K = \P^{-1} \B \P$.  Next,
define the time evolution operator  $\U(t, t_0) = \exp[\D (t -
t_0)]$ and use the method of variation of constants to
rewrite the linearized equation of motion as the following
integral equation~\cite{fv}
\begin{equation}
\label{int_eq}
\z(t) = \U(t, t_0) \z(t_0) + \int_{t_0}^t \U(t,s) \K(s) \z(s) \,
ds,
\end{equation}
where $\K(s) \equiv \K[\x(s); s]$ denotes \K\ evaluated along the
driving trajectory.

The linear stability of synchronization to the driving trajectory,
$\x(t)$, is determined by the behavior of $\| \z(t) \|$ in the 
$t \rightarrow \infty$ limit.  To determine this behavior
the remainder of the calculations in this section closely follow
those of Chapter~6 in Ref.~\cite{fv}.  Begin by using norms to
obtain the following inequality~\cite{gv}
\begin{displaymath}
\| \z(t) \| \leq \| \U(t,t_0) \| \: \| \z(t_0) \| 
+ \int_{t_0}^t \| \: \U(t,s) \| \: \| \K(s) \| \: \| 
\z(s) \| ds.
\end{displaymath}
It is known that if $\Re[\Lambda_1] < 0$ then $\| \U(t,t_0) \|
\leq C \exp[- \mu (t-t_0)]$ for appropriate choices of $C$ and
$\mu$. Inserting this into the inequality and  
applying Gronwall's Theorem~\cite{fv} yields
\begin{displaymath}
\| \z(t) \| \leq C \: \| \z(t_0) \| \exp \left[ \int_{t_0}^t
\left( C \: \| \K(s) \| -  \mu \right) ds \right].
\end{displaymath}
Hence, a sufficient condition for linearly stable 
synchronization is
\begin{equation}
\label{norm}
\mu > \lim_{t \rightarrow \infty} \frac{C}{t - t_0} \int_{t_0}^t 
\| \K(s) \| \: ds ,
\end{equation}
which depends on $C$, $\mu$, $\| \K \|$, and the measure of the
driving signal, $\x(t)$.  

The constants $C$ and
$\mu$ depend explicitly on the choice of matrix norm.  A
common choice is the Frobenius norm~\cite{gv}
\begin{displaymath}
\| \M \| \equiv \left[ \Tr \left( \M^{\dagger} \M \right)
\right]^{1/2} = \left[ \sum_{\alpha, \: \beta=1}^d \left| 
M_{\alpha \beta} \right|^2 \right]^{1/2}  ,
\end{displaymath}
where the $\dagger$ denotes Hermitian conjugate and Greek
subscripts denote elements of matrices ($M_{\alpha \beta}$
is the $\alpha$ $\beta$ element of the matrix \M).  Applying this
norm to $\U(t,t_0)$ and using the rank ordering of the
$\Lambda$'s implies that, in the large $t$ limit, $\|
\U(t,t_0) \| \simeq \exp[\Re[\Lambda_1] (t-t_0)]$, an
approximation that becomes exact in the $t \rightarrow \infty$
limit.  Therefore, in this limit, $C=1$ and $\mu = 
-\Re[\Lambda_1]$.

Inserting these results into Eq.~(\ref{norm}) indicates that the 
sufficient condition for linear stability of synchronization 
along the driving trajectory $\x(t)$ is
\begin{equation}
\label{cond1}
- \Re[\Lambda_1] > \lim_{t \rightarrow \infty} \frac{1}{t - t_0}
\int_{t_0}^t \| \K [\x(s); s] \| \: ds .
\end{equation}
The fact that the right hand side of Eq.~(\ref{cond1}) is
positive semi-definite implies that $\Re[ \Lambda_1]$ must be
negative for there to be any chance of satisfying the rigorous
condition.

Because the decomposition given by Eq.~(\ref{decomp}) is not 
unique this condition for linear stability of synchronization, 
Eq.~(\ref{cond1}), requires to an examination of \K.   Notice that  
$\| \K \| = \| \widetilde{\DF} - \widetilde{\Q} \|$ where 
$\widetilde{\DF} \equiv \P^{-1} \DF \P$ and $\widetilde{\Q}
\equiv  \P^{-1} [\A + \DE(\zero)] \P$.  Thus, the right hand side
of Eq.~(\ref{cond1}) can be rewritten as
\begin{displaymath}
I = \lim_{t \rightarrow \infty} \frac{1}{t-t_0} \int_{t_0}^t
\left[ \sum_{\alpha, \: \beta=1}^d \left( 
\widetilde{DF}_{\alpha \beta} [\x(s); s] - \widetilde{Q}_{\alpha
\beta} \right)^2 \right]^{1/2} ds,
\end{displaymath}
where the ambiguity of the decomposition is confined to the matrix
$\widetilde{\Q}$.  The role of $I$ in 
Eq.~(\ref{cond1}) leads us to conjecture that minimizing $I$
produces the best chance for synchronization.  We make this conjecture
despite the fact that $\Re[\Lambda_1]$ depends on this decomposition.
Hence, we define the optimal value of $\Q=\A + \DE(\zero)$ as the one
which minimizes $I$.

It is straightforward, by setting $\partial I / \partial 
\widetilde{\Q} = \zero$, to show that the optimal value for \Q\
is given by
\begin{displaymath}
\Q = \lim_{t \rightarrow \infty} \frac{1}{t - t_0} \int_{t_0}^t
\DF[\x(s); s] \: ds \equiv \left\langle \DF \right\rangle,
\end{displaymath}
where $\left\langle \bullet \right\rangle$ denotes a time
average on the invariant measure $\x(s)$.  This result implies 
that the optimal decomposition in Eq.~(\ref{decomp}) is
\begin{eqnarray}
\label{def_A}
\A & \equiv & \left\langle \DF \right\rangle - \DE(\zero) \\
\label{def_B}
\B(\x ;t) & \equiv & \DF(\x; t) - \left\langle \DF \right\rangle,
\end{eqnarray}
and the condition for linear stability of the invariant trajectory
in the synchronization manifold is
\begin{equation}
\label{cond2}
- \Re[\Lambda_1] > \left\langle \| \P^{-1} \left[ \B(\x ;t) 
\right] \P \| \right\rangle .
\end{equation}

Equations~(\ref{def_A})--(\ref{cond2}) are the major results of
this section.  Together they represent definitions and a criterion
which indicate when synchronous motion along a particular driving
trajectory is guaranteed to be stable to small perturbations in 
directions transverse to the synchronization manifold.   As 
indicated by our examples and Appendix~\ref{geometry2} this 
criterion can be used to design couplings that are guaranteed to
result in stable synchronization.  Unfortunately, the criterion
is only sufficient, not necessary {\em and} sufficient.  Thus,
one can expect, and our numerical experiments shown, that it is 
possible for a coupling scheme to fail this criterion and still
produce stable synchronization.  This is due, in part, to the
fact that the derivation of Eq.~(\ref{cond2}) involved 
inequalities of norms which will tend to overestimate the
necessary coupling strengths.  The relationship between these
results and control theory are discussed below in 
Appendix~\ref{control}.

As defined above, the decomposition in Eqs.~(\ref{def_A}) and 
(\ref{def_B}) is optimal in the sense that it minimizes the right 
hand side of Eq.~(\ref{cond2}).  We have conjectured that this
gives one the best chance at satisfying the inequality.
To support this conjecture we have the following circumstantial
evidence.  If we insert Eq.~(\ref{def_B}) into a Volterra expansion
of Eq.~(\ref{int_eq}) then, to second order, the
criteria for linear stability is $\Re[\Lambda_1] < 0$ (see
the next section).  For any other decomposition this simple
approximate stability criteria is only correct to first order.
If the driving trajectory is a fixed point then $\B(\x ;t) = \zero$
and Eq.~(\ref{cond2}) reduces to $\Re[\Lambda_1] < 0$.  This result
is what one would expect from a simple linear stability.  However,
it does not occur for other decompositions.  To address the full
optimality issue one must compare the size of $\left\langle \|
\P^{-1} \B \P \| \right\rangle$ to the size of the first eigenvalue
of \A.  We are unable to make this comparison due to the complicated
dependence of these eigenvalues on the coupling strengths, $\DE(
\zero)$.  (The idea of using a time average to define \A\ is also
suggested in Ref.~\cite{pddy}.)

Because the stability criterion depends explicitly on  the
driving trajectory the measure associated with the dynamics is
of crucial importance.  An early paper by Gupte and
Amaritkar~\cite{ga} used unstable periodic orbits as driving
trajectories for synchronization.  They (and others) found
that, for fixed coupling strength, synchronization is stable
for some driving trajectories and unstable for other driving
trajectories~\cite{gb,hcp2}.  In order to determine a coupling
strength that results in a stable synchronization manifold (i.e.,
synchronous motion is stable on all possible driving trajectories)
notice that the right hand side of Eq.~(\ref{cond2}) is a time 
average. 

Recently, Hunt and Ott~\cite{ho} examined time averages of
functions for different measures of the dynamics of a chaotic 
system.  They found that, for chaotic systems, time averages tend
to take on their largest values for measures confined to the 
unstable periodic orbits with the shortest periods.  (This
effect was also observed in Ref.~\cite{abs}.)  They also
found that if an unstable periodic orbit with a somewhat higher
period has the largest time average then this value is 
usually only a small increase compared to orbits with
lower periods.  Therefore, if synchronization is stable to fixed
points and periodic orbits (with short periods) located in the 
chaotic attractor, one may consider this as an indication that 
synchronized chaotic motion is also stable (an assumption also
proposed in Ref.~\cite{gb}).  This assumption 
is supported by the Hunt and Ott who claim that a nonperiodic
orbit will yield the maximum time average on a set of measure
zero in the parameter space associated with the parameters of
\F.  Finally, this paper presents a crude formula for determining
how high in period one must look in order to insure that the
largest time averages have been found.  (The issue of obtaining
obtain time averages for dynamics on chaotic attractors by
weighting time averages on the unstable periodic orbits has
been addressed by Cvitanovi\'c~\cite{cvit}.)

In general the stability condition given by Eq.~(\ref{cond2})
requires that one integrate the norm of a matrix along the
driving trajectory.  As such these equations could be difficult
to use as an explicit test to determine whether a given coupling 
scheme will produce synchronization.  (Despite this they are still
far more efficient than the Lyapunov exponent approximations found
in most other papers.)  However, for simple systems, many of the
important calculations can be performed analytically (see below
Appendix~\ref{lorenz} and Ref.~\cite{wm} for examples). 
Therefore, the criterion of Eqs.~(\ref{def_A})--(\ref{cond2}),
along with the results of Ref.~\cite{ho} may well allow one to
determine coupling schemes that insure linear stability of the
entire synchronization manifold.

\section{Theory Part II:  Approximate Results}
\label{theory2}
Even with the rigorous results of the previous section one often
desires a ``quick and dirty'' criterion to approximate whether a
particular coupling scheme and/or coupling strength will produce
linearly stable synchronous motion.  The major result of this section
is just such a criterion.

Begin by noting that an infinite series solution to Eq.~(\ref{int_eq}) 
can be obtained by inserting this equation into itself to form a 
Volterra expansion~\cite{nrc}
\begin{displaymath}
\z(t) = \left[ \U(t, t_0) + \sum_{j=1}^\infty \M^{(j)} (t, t_0)
\right] \z(t_0),
\end{displaymath}
where
\begin{eqnarray*}
\lefteqn{\M^{(j)}(t, t_0) = \int_{t_0}^t ds_1\int_{t_0}^{s_1} ds_2 
\cdots \int_{t_0}^{s_{j-1}} ds_j } \\
& \times & \left[ \U(t, s_1) \K(s_1) \U(s_1, s_2)
\K(s_2)  \cdots \K(s_j) \U(s_j, t_0) \right] .
\end{eqnarray*}

Truncating this solution at $j=1$ yields the following approximate
solution to Eq.~(\ref{int_eq})
\begin{equation}
\label{approx}
\z(t) \simeq \left[ \U(t, t_0) + \M^{(1)}(t, t_0) \right]
\z(t_0).
\end{equation}
(Provided the series is well behaved, it converges, etc., one can
obtain better solutions than the one discussed below by retaining
more terms in the series.)  To this approximation if all elements
of $\U + \M^{(1)}$ decay in time then $\lim_{t \rightarrow 0} \| \z(t)
\| = 0$, and synchronization is linearly stable.

It is useful to divide the discussion of $\M^{(1)}$ into an analysis
of diagonal and off diagonal elements.
\begin{enumerate}
\item (Diagonal elements)  \\
For these elements
\begin{eqnarray*}
\lefteqn{\left[ M^{(1)} \right]_{\alpha \alpha} = \exp \left[ 
\Lambda_\alpha(t-t_0) \right] \sum_{\mu, \: \nu=1}^d 
P^{-1}_{\alpha \mu} } \\
 & \times & \left[ \left( \int_{t_0}^t  DF_{\mu \nu} 
(s) ds \right)  - \left\langle DF_{\mu \nu} \right\rangle
(t-t_0) \right] P_{\nu \alpha},
\end{eqnarray*}
were the optimal decomposition of Eqs.~(\ref{def_A}) and
(\ref{def_B}) have been used.  For this decomposition the term in
large square brackets vanishes in the large $t$ limit. 
For any other decomposition this term can be expected to grow at
least linearly with time.  Depending on the strength of growth
this term could offer significant competition to the anticipated
exponential decay from $\exp\left[ \Lambda_\alpha(t - t_0)
\right]$.

\item (Off diagonal elements)  \\
For these elements
\begin{eqnarray*}
\left[ M^{(1)} \right]_{\alpha \beta} & = & \exp \left(
\Lambda_\alpha  t - \Lambda_\beta t_0 \right) \\
 & \times & \int_{t_0}^t  \exp
\left[ (\Lambda_\beta - \Lambda_\alpha) s \right] K_{\alpha
\beta} (s) \: ds.
\end{eqnarray*}
For the dynamical systems we are examining the driving signal, \x, is 
bounded and the Jacobian is bounded (typically \x\ is confined to a
compact attractor).  Therefore,
\begin{displaymath}
\left[ M^{(1)} \right]_{\alpha \beta} \leq \frac{K_{\alpha
\beta}^{(\max)}} {\Lambda_\beta - \Lambda_\alpha} \left[ \exp
\left( \Lambda_\beta t  \right) - \exp \left( \Lambda_\alpha t
\right) \right],
\end{displaymath}
where $K_{\alpha \beta}^{(\max)}$ is the maximum value taken by
$K_{\alpha \beta}(s)$ on the driving trajectory and we have assumed
the eigenvalues of \A\ are distinct.  
\end{enumerate}

This analysis of $\M^{(1)}$ indicates that if $\Re[\Lambda_1] <
0$ then the off diagonal elements of the matrix $\U + \M^{(1)}$
die off exponentially for $t > t_0$, while the diagonal elements 
die off at a rate controlled by the convergence of the time average
$\left\langle \DF \right\rangle$.  Therefore, to second order we have 
the following simple condition for linear stability of a trajectory
on the synchronization manifold to transverse perturbations
\begin{equation}
\label{cond3}
\Re \left[ \Lambda_1 \right] < 0.
\end{equation}
Equation~(\ref{cond3}) is the major results of this section.  It,
and Eq.~(\ref{def_A}), represent a ``quick and dirty'' criterion to
determine  whether or not synchronization will occur for a
particular coupling scheme.

Equation~(\ref{cond3}) is completely consistent with
the rigorous results of the previous section.  As expected the
approximate criterion given by Eqs.~(\ref{def_A}) and (\ref{cond3}) depends
explicitly on the measure of the driving trajectory.  
An appealing aspect of Eq.~(\ref{cond3}) is
that the analytic and computational burden associated with computing \A\ and
it eigenvalues is relatively minor.  Thus, one can quickly search over
many types of coupling schemes.

Our numerical experiments show that Eq.~(\ref{cond3}) is actually
a good approximation to the {\em minimum} coupling needed to achieve
linearly stable synchronization.  One possible explanation for the 
success of this approximation is a ``folk theorem'' from control 
theory discussed by Brogan~\cite{brogan}.  ``A commonly used stability
analysis technique is the so-called frozen coefficient method, in which 
all time-varying coefficients are frozen and then the system stability
is analyzed as if it were a constant coefficient system''~\cite{brogan}.  
Brogan goes on to claim that ``when used with caution, this approach
will {\em usually} give the correct result''.  As a final comment we
note that Ref.~\cite{pddy} also examined \A\ to determine the linear
stability of synchronization.

\section{Numerical Experiments}
\label{results}
This section presents results from numerical experiments we have
performed to test the conditions derived in
Sections~\ref{theory1} and \ref{theory2}.  The results from
the first example are mostly numerical and focus on the
approximate condition.  By the last example almost all of the
calculations are performed analytically and focus on both the
rigorous and the approximate condition (see Appendix~\ref{other} 
for yet another example).   When appropriate we
will use $\epsilon^{(c)}$ to denote critical coupling strengths
where the invariant trajectory on the synchronization 
manifold becomes linearly stable.

\subsection{Rossler Example}
The Rossler system is the following set of three coupled ODE's
\begin{eqnarray*}
\frac{d x}{dt} & = & -y - z \\
\frac{d y}{dt} & = & x + a y \\
\frac{d z}{dt} & = & b + z(x - c) ,
\end{eqnarray*}
where $a=0.2$, $b=0.2$ and $c=9$.  For these parameter settings 
the dynamics of this system has a chaotic attractor, and two 
unstable fixed points located at $\x_\pm = (-a y_\pm, y_\pm, -y_\pm)$
where
\begin{displaymath}
y_\pm = - \frac{c}{2a} \left[ 1 \pm \left( 1 - \frac{4ab}{c^2} 
\right)^{1/2} \right].
\end{displaymath}
Only the fixed point at $\x_{-}$ is near the attractor.

To simplify matters we 
use diagonal coupling, $\E(\x - \y) = \text{diag} (\epsilon_1, \epsilon_2,
\epsilon_3) \cdot (\x - \y)$.  For this type of coupling it is
easy to show that Eq.~(\ref{def_A}) leads to
\begin{displaymath}
\A = \left[
\begin{array}{ccc}
 - \epsilon_1 & -1 & -1 \\
 1 & a - \epsilon_2 & 0 \\
 \left\langle z \right\rangle & 0 & \left\langle x \right\rangle
- c - \epsilon_3
\end{array}
\right].
\end{displaymath}

\begin{figure}
\vspace{0.0in}
\begin{center}
\leavevmode
\hbox{%
\epsfxsize=3.375in
\epsffile{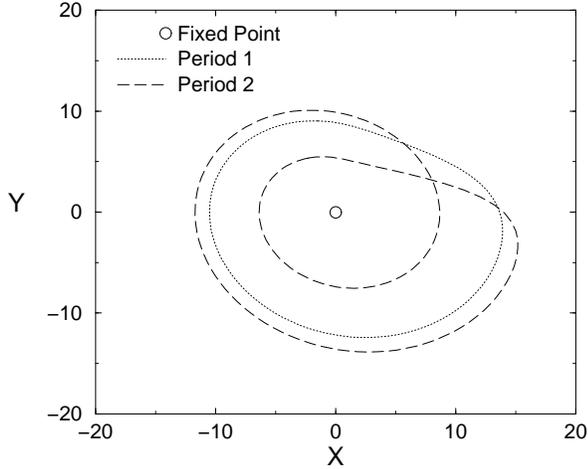}}
\end{center}
\caption{Fixed point, period~1 and period~2 orbits of the Rossler
systems.
\label{fig.Ros.orbits}}
\end{figure}

Our numerical experiments examine four different measures of the 
driving dynamics.  One is the natural
or Sinai-Bowen-Ruelle (SBR) measure on the chaotic attractor.  
This arises from following an arbitrary trajectory on the
chaotic attractor.  The others are
Dirac measures associated with the $\x_{-}$ fixed point, and
unstable period~1 and period~2 orbits. 
The trajectories associated with
the fixed point and the periodic orbits are shown in
Fig.~\ref{fig.Ros.orbits}.

Because the characteristic equation for \A\ is a cubic that can not 
be easily factored for arbitrary values of the $\epsilon$'s
most of the results were obtained
numerically.  (We have calculated these eigenvalues using a symbolic
manipulator to perform the necessary algebra.  The expressions
are not particularly enlightening, and will not be
presented.)

\subsubsection{Fixed point measure}
\begin{figure}
\vspace{0.0in}
\begin{center}
\leavevmode
\hbox{%
\epsfxsize=3.375in
\epsffile{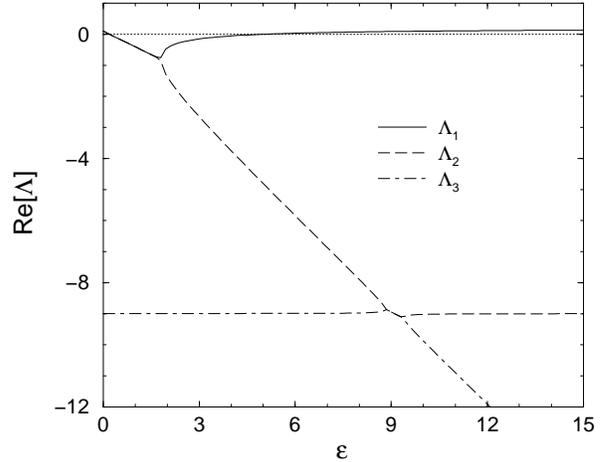}}
\end{center}
\caption{The real parts of the eigenvalues of \A\ as functions of
$\epsilon$ for $x$-driving.  \A\ comes from the Rossler system and 
the driving trajectory is the fixed point $\x_{-}$.  The dashed line
at $\Lambda=0$ is inserted as a visual aid.
\label{fig.Ros.eig.0}}
\end{figure}

On this measure $\B(\x ;t) = \zero$, so the rigorous and 
approximated criterion are both $\Re[\Lambda_1] < 0$ (a familiar
result from stability analysis of fixed points found in text books). 
Figure~\ref{fig.Ros.eig.0} shows the real parts of the eigenvalues
of \A\ as functions of $\epsilon \equiv \epsilon_1$ when the driving
trajectory is $\x_{-}$ and we only use the first component of $\x_{-}$
as the driving signal ($\epsilon_2 = \epsilon_3 = 0$).  This type of
driving is typically called $x$-driving.  (A dotted line at $\Lambda = 
0$ is inserted as a visual aid.)  The figure indicates that, initially,
$\Lambda_1$ and $\Lambda_2$ are complex conjugate pairs with positive
real part.  However, as $\epsilon$ increases the real parts become negative 
at $\epsilon^{(c)} = 0.19750\ldots$.  As $\epsilon$ continues to increase
$\Lambda_1$ and $\Lambda_2$ become real at $\epsilon \simeq 1.8$, with
$\Lambda_1$ increasing and $\Lambda_2$ decreasing.  $\Lambda_1$ eventually
becomes positive
at $\epsilon = 4.9975\ldots$.   The figure also indicates
that $\Lambda_2$ and $\Lambda_3$ merge into complex
conjugate pairs but soon split back into pure reals at $\epsilon \simeq
9$.  If the second component of $\x_{-}$ is used as the driving
signal ($y$-driving) then a similar figure is produced, however 
$\Lambda_1$ remains negative for large $\epsilon$.

The interesting feature of Fig.~\ref{fig.Ros.eig.0} is the prediction 
of synchronous behavior for $x$-driving when $\epsilon$ is in the
range between about 0.2 and about 5.
Although no numerical experiment can prove the correct values of
$\epsilon^{(c)}$, our experiments
indicates that when $x$-driving is used a bifurcation from
unstable (stable) synchronization to a stable (unstable)
synchronization appears to occur at $\epsilon^{(c)} =
0.19750\ldots$ ($4.9975\ldots$).

The results of all of our numerical experiments on the Rossler
system appear in Table~\ref{table.ross}.  The table indicate
that for both $x$ and $y$-driving synchronization to $\x_{-}$
begins at $\epsilon^{(c)} \simeq 0.2$.  For $x$-driving
synchronization ends at $\epsilon^{(c)} \simeq 4.9$, while for
$y$-driving it persists for arbitrarily large values of
$\epsilon$.
When $z$-driving is used $\Lambda_1$ and $\Lambda_2$ are complex 
conjugate pairs with positive real parts throughout the entire 
range of $\epsilon$.  Thus, the rigorous and approximate
condition predict that synchronization will not  occur at
$\x_{-}$ for this type of driving.  The numerical experiments
appear to verified this result.  (We remark that the rigorous
condition predicts, and we were able to verify, a range of values
for $\epsilon$ which will result in synchronization about
$\x_{+}$ for $z$-driving.)

\subsubsection{Periodic orbit measures}
On these measures (as well as the SBR measure) $\B(\x ;t) \neq
\zero$, so the rigorous and approximate
criterion are not the same.  We
will only consider the approximate condition and save a
discussion of the rigorous condition for later examples.

The matrix \A\ is a function of time averages over the driving
trajectory.  Table~\ref{table.averages} shows numerically determined values
for $\left\langle x \right\rangle$, $\left\langle y \right\rangle$,
and $\left\langle z \right\rangle$ on each of our measures.

Numerical experiments using the measure associated with periodic
orbits indicate that the eigenvalues of \A\ undergo the
same type of splittings and mergings found for the fixed point 
case.  More importantly for the period~1 (2) orbit
$\Re[\Lambda_1]$ becomes negative at $\epsilon
\simeq 0.04$ (0.07) for both $x$ and $y$-driving.  In addition,
for $x$-driving and the period~1 (2) orbit, 
$\Re[\Lambda_1]$ becomes positive at $\epsilon \simeq 4.84$
(4.87).  When $y$-driving is used $\Lambda_1$ remains negative
for large $\epsilon$.  The approximate criteria predicts a
change in stability each time $\Re[\Lambda_1]$ changes sign.

\begin{figure}
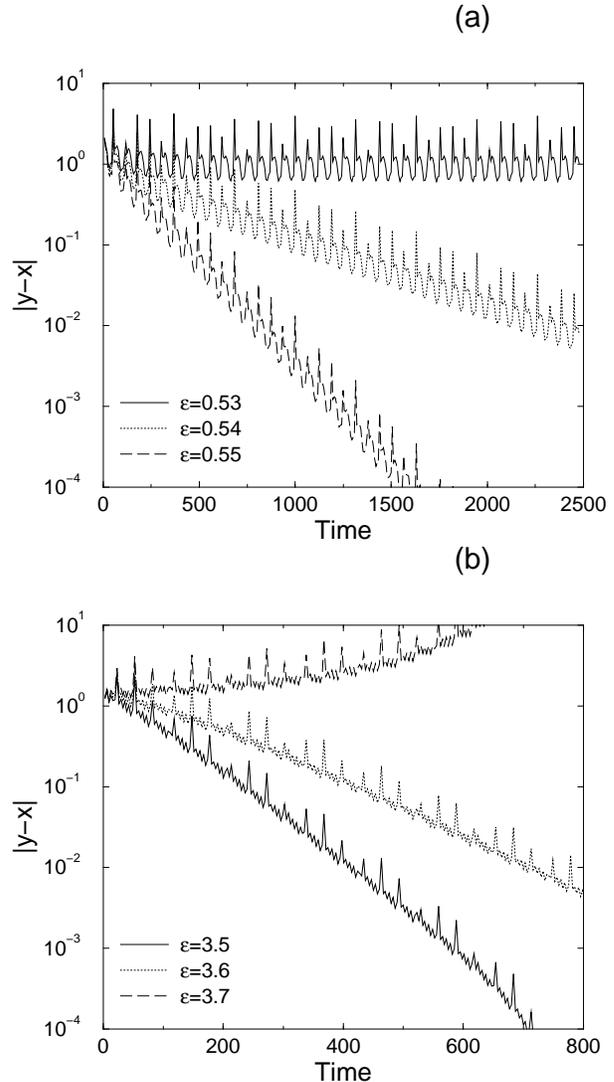

\vspace{0.0in}
\begin{center}
\leavevmode
\hbox{%
\epsfxsize=3.375in
\epsffile{fig.Ros.1.x_drive.a}}
\hbox{%
\epsfxsize=3.375in
\epsffile{fig.Ros.1.x_drive.b}}
\end{center}
\caption{Results of numerical test for critical coupling
strengths. The Rossler system synchronized to the period~1 
orbit.  The period of the orbit is approximately six units
of time and $x$-driving is used.  (a) output is every 
106/96th of the period.  (b) output is every 53/96th of 
the period.
\label{fig.Ros.1.drive}}
\end{figure}

The results of numerical experiments to test
these predictions are shown in Figs.~\ref{fig.Ros.1.drive}. 
The figures indicate that if $x$-driving is used and $\epsilon
= 0.53$ then the response system remains a finite distance 
from the period~1 orbit.  However, if $\epsilon = 0.54$ then 
the driven dynamics exponentially converges onto the period~1
orbit.  The figures also show that if $\epsilon = 3.8$ then
the response system appears to diverge while it exponentially
converges to the period~1 orbit for $\epsilon = 3.7$. 

As shown in Table~\ref{table.ross}, numerical tests of these 
predictions indicate that the lower (higher) critical coupling 
strength is higher (lower) than that predicted by the approximate
criterion.  Thus, the range of coupling strengths for which 
synchronization occurs is slightly smaller than that predicted 
by the approximate criterion.
This is to be expected since the rigorous criterion implies that
it is not enough that $\Re[\Lambda_1]$ be negative, it must be {\em
sufficiently} negative to overcome the fluctuations represented by
$\left\langle \| \P^{-1} \B \P \| \right\rangle$.  However, given
the ``quick and dirty'' nature of the approximate criterion we
believe that it yields reasonable predictions for coupling
strengths that produce synchronization.

\subsubsection{SBR measure}
\begin{figure}
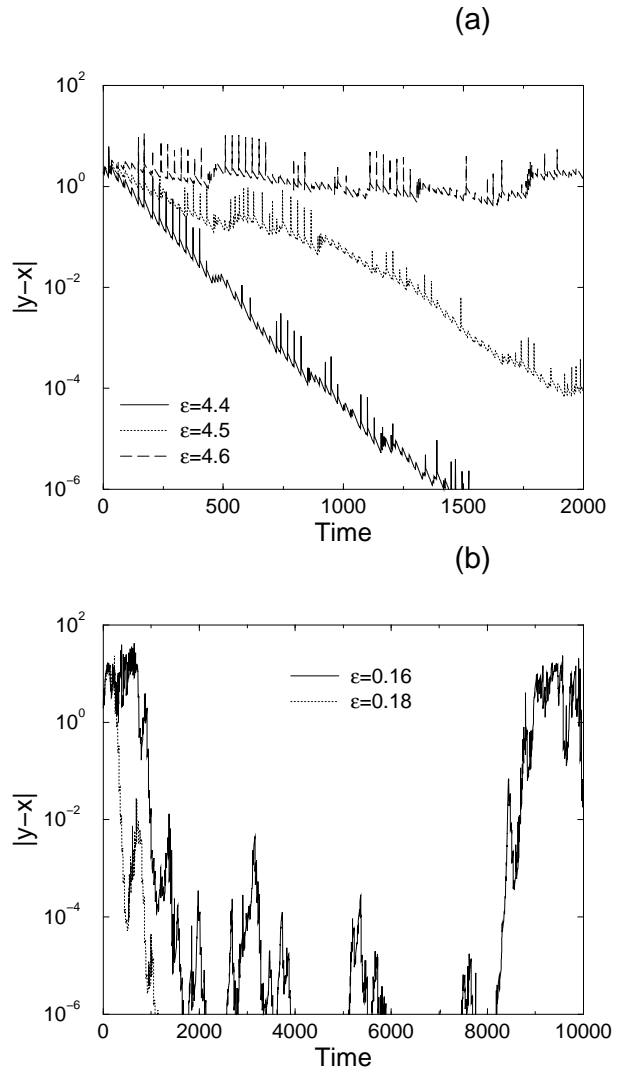

\vspace{0.0in}
\begin{center}
\leavevmode
\hbox{%
\epsfxsize=3.375in
\epsffile{fig.Ros.SBR.x_drive}}
\hbox{%
\epsfxsize=3.375in
\epsffile{fig.Ros.SBR.y_drive}}
\end{center}
\caption{Results of numerical test for critical coupling
strengths. The Rossler system synchronized to a chaotic
trajectory.  (a) $x$-driving with output every one unit of
time.  (b) $y$-driving with output every two units of time.
\label{fig.Ros.SBR.drive}}
\end{figure}

The results of tests conducted on the SBR measure are much more 
complicated that those obtained on the simple measures examined
above. Sample plots of $\| \y - \x \|$ for various values of
$\epsilon$ and  types of driving are shown in
Figs.~\ref{fig.Ros.SBR.drive}.  These  figures indicate the
bursting phenomenal observed by previous researchers when
discussing synchronization.  The degree to which the  response
system is ``synchronized'' to the drive system clearly depends on
the degree to which one is willing to tolerate bursts. 

For example, when $y$-driving is used
Fig.~\ref{fig.Ros.SBR.drive}b indicates that $\| \y - \x \| \sim
10^{-4}$ can occur for as long as a thousand times around the
attractor before climbing to $\| \y - \x \| \sim 1$.  (For
$\epsilon = 0.15$ we have observed $\| \y - \x \| < 10^{-4}$ for
as many as 4000 times around the attractor.)  These examples
confirm that one must be cautious about discussing the condition
for linear stability of synchronization when the SBR measure is
used.  The values shown in Table~\ref{table.ross} were selected
because $\| \y - \x \|$ steadily decayed to a small value and did
not significantly increase  for the next 50,000 time steps.  In
addition we required that $\| \y - \x \|$ exhibit the same
behavior for values of $\epsilon$ close to $\epsilon^{(c)}$ and
on the stable side of the threshold.

However, because SBR 
trajectories are dense they will eventually come arbitrarily close
to any unstable region of the attractor.  The values of
$\epsilon^{(c)}$ listed in Table~\ref{table.ross} for the SBR
measure are outside the range listed for the period~1 and period~2
orbits.  Therefore, if the SBR orbit comes sufficiently close to
either of these orbits and/or stays close for a sufficiently long 
time then a burst could happen~\cite{hcp2}.  The results implied 
by Table~\ref{table.ross} indicate for our numerical experiments
the SBR trajectory did not come sufficiently close the either 
the period~1 or period~2 orbits.

\subsection{Lorenz Example}
The Lorenz system is the following set of three coupled ODE's
\begin{eqnarray*}
\frac{d x}{dt} & = & \sigma (y - x) \\
\frac{d y}{dt} & = & rx - y - xz \\
\frac{d z}{dt} & = & xy - bz ,
\end{eqnarray*}
where $\sigma=10$, $b=8/3$, and $r=60$.  For these parameter values
the dynamics of the system has a chaotic attractor and three unstable
fixed points, one of which is at the origin.
\begin{figure}
\begin{center}
\leavevmode
\hbox{%
\epsfxsize=3.375in
\epsffile{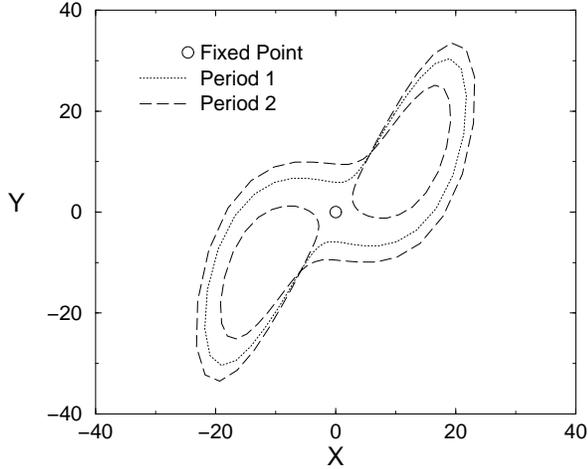}}
\end{center}
\caption{Fixed point, period~1 and period~2 orbits of the Lorenz
system.
\label{fig.Lor.orbits}}
\end{figure}

This example is particularly instructive because if the coupling
is given by 
\begin{displaymath}
\DE(\zero) = \left[
\begin{array}{ccc}
\epsilon_1 & \epsilon_4 & 0 \\
\epsilon_3 & \epsilon_2 & 0 \\
0 & 0 & \epsilon_z
\end{array} 
\right]
\end{displaymath}
then many of the calculations can be performed analytically. 
Furthermore, this type of coupling allows the $x$ {\em or} $y$
variable to drive both the $x$ {\em and} $y$ equations.  Ott and
Ding~\cite{od} have shown that this type of coupling is useful
when not all variables are measurable.  Their work, and the result
presented below, indicate that this type of driving may produce or
guarantee synchronization when purely diagonal coupling does not.

For this dynamical system it is easy to show that
Eq.~(\ref{def_A}) leads to
\begin{displaymath}
\A = \left[
\begin{array}{ccc}
 -\sigma - \epsilon_1 & \sigma - \epsilon_4 & 0 \\
 r - \left\langle z \right\rangle - \epsilon_3 & -1 - \epsilon_2 
 & - \left\langle x \right\rangle \\
 \left\langle y \right\rangle & \left\langle x \right\rangle & -b
 - \epsilon_z
\end{array}
\right].
\end{displaymath}

We examine the SBR measure and 
Dirac measures associated with the fixed point at the origin, a
period~1, and a period~2 orbit. The trajectories associated with 
these measures are shown in Fig.~\ref{fig.Lor.orbits}.  
Table~\ref{table.averages} shows numerically calculated values for
$\left\langle x \right\rangle$, $\left\langle y \right\rangle$, and
$\left\langle z \right\rangle$, on each of these trajectories.
The eigenvalues of \A\ are 
\begin{eqnarray}
\Lambda_z & = & -b - \epsilon_z \nonumber \\
\Lambda_\pm & = & \frac{- (\sigma + 1 + \epsilon_1 +
\epsilon_2)}{2} \nonumber \\
\label{lor_lambda_pm}
 & \pm & \frac{1}{2} \left( \left[ (\sigma + \epsilon_1)
- (1 + \epsilon_2) \right]^2 \right. \nonumber \\
 & + & \left. 4(\sigma - \epsilon_4) (r - 
\left\langle z  \right\rangle - \epsilon_3) \right)^{1/2},
\end{eqnarray}
and, in practice, $\Lambda_1$ could be either $\Lambda_{+}$ or 
$\Lambda_z$.

\subsubsection{Fixed point measures}
On these measures $\B(\x ;t) = \zero$, and the rigorous and
approximate criterion are both $\Re[\Lambda_1] < 0$.  Because,
$\Lambda_z = -b - \epsilon_z < 0$ for all positive values of 
$\epsilon_z$ we focus our attention on $\Lambda_\pm$.
Analysis of Eq.~(\ref{lor_lambda_pm}) indicates that if diagonal
$x$-driving is used ($\epsilon_2 = \epsilon_3 = \epsilon_4 = 0$)
then the critical coupling strength associated
with $\Re[\Lambda_1] =0$ is $\epsilon_1^{(c)} = \sigma(r-1) = 590$.
If diagonal $y$-driving is used ($\epsilon_1 = \epsilon_3 = 
\epsilon_4 = 0$) then the critical coupling strength
is $\epsilon_2^{(c)} = r-1 = 59$.
Finally, it is easy to shown that if diagonal $z$-driving is used
then $\epsilon_1 = \epsilon_2 = \epsilon_3 = \epsilon_4 = 0$ and 
$\Re[\Lambda_1] > 0$.  Therefore, for $z$-driving the rigorous
criterion can not be satisfied.
The results of all of our numerical experiments on
the Lorenz system appear in Table~\ref{table.lor}.

\subsubsection{Periodic orbit measures}
\begin{figure}
\begin{center}
\leavevmode
\hbox{%
\epsfxsize=3.375in
\epsffile{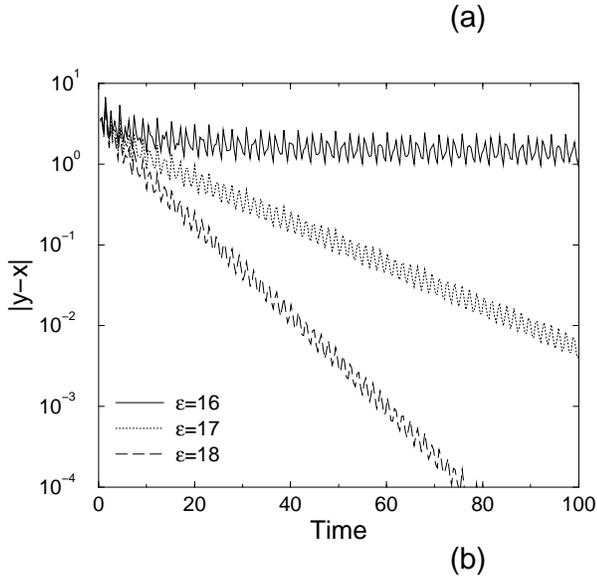}}
\hbox{%
\epsfxsize=3.375in
\epsffile{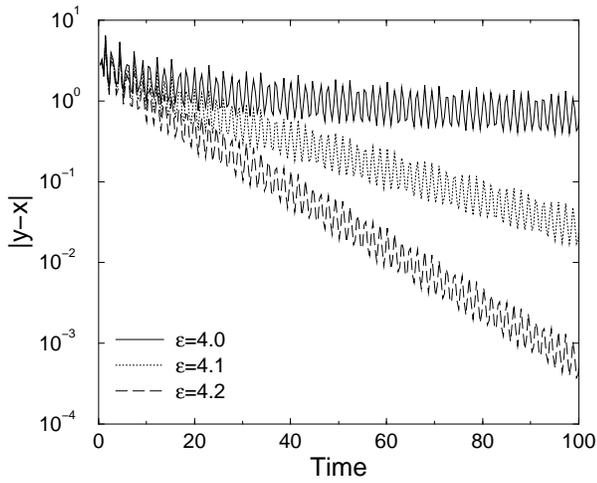}}
\end{center}
\caption{Results of numerical test for critical coupling
strengths. The Lorenz system synchronized to a period~1
orbit.  The period of the orbit is approximately one unit
of time and output is every 19/50th of the period. (a) 
$x$-driving. (b) $y$-driving.
\label{fig.Lor.1.drive}}
\end{figure}

On these measures (as well as the SBR measure) $\B(\x ;t)
\neq \zero$.  An analysis of the approximate condition of 
Eq.~(\ref{cond3}), in conjunction with Eq.~(\ref{lor_lambda_pm}),
indicates that the approximation to the critical coupling strength 
is $\epsilon_1 = \sigma[r - \left\langle z \right\rangle -1] 
\simeq 42$ for diagonal $x$-driving, and $\epsilon_2 = r - 
\left\langle z \right\rangle -1 \simeq 4.2$ for diagonal $y$-driving.
Numerical experiments on the period~1 orbit are shown in 
Fig~\ref{fig.Lor.1.drive}.  They indicate that $\epsilon_1^{(c)}
\simeq 17$ for
$x$-driving and $\epsilon_2^{(c)} \simeq 4.2$ for $y$-driving.
Although the error in our approximation for $x$-driving may appear
large, comparing it to the value $\epsilon^{(c)} = 590$
needed for the fixed point at the origin indicates that it may
not be very large.

\subsubsection{SBR measure}
\begin{figure}
\begin{center}
\leavevmode
\hbox{%
\epsfxsize=3.375in
\epsffile{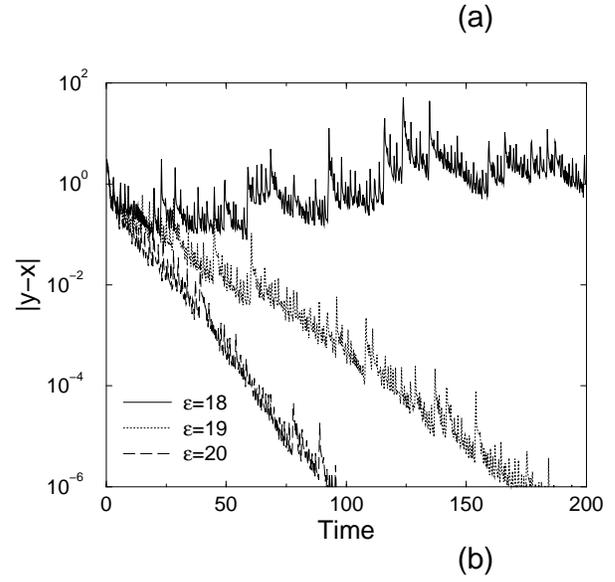}}
\hbox{%
\epsfxsize=3.375in
\epsffile{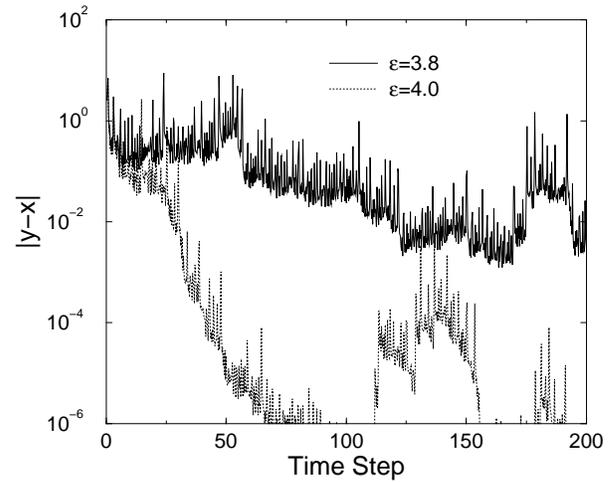}}
\end{center}
\caption{Results of numerical test for critical coupling
strengths. The Lorenz system synchronized to a chaotic
trajectory.  The output is every 1/4 units of time.  (a) 
$x$-driving.  (b) $y$-driving.
\label{fig.Lor.SBR.drive}}
\end{figure}

Results of the tests conducted on the SBR
measure are much more complicated than those obtained on the  simple
measures examined above.  Figures~\ref{fig.Lor.SBR.drive} exhibit the 
bursting behavior observed when the trajectory of the response system
approaches a region of phase space that is unstable to perturbations 
perpendicular to the synchronization manifold.  Because SBR trajectories
are dense they will eventually encounter any such region on the attractor.
Thus, although Fig.~\ref{fig.Lor.SBR.drive}a appears to indicate that
the system will synchronize when $\epsilon_1 > 18$ for $x$-driving
we know that the fixed point at the origin is unstable
for this value of $\epsilon_1$.  Therefore, if the trajectory
comes sufficiently close to $\x = \zero$ then a burst will occur. 

\subsection{Ott--Sommerer example}
The third dynamical system we examine is the Ott--Sommerer~\cite{os}
set of four nonautonimous ODE's
\begin{eqnarray}
\frac{d x}{dt} & = & v_x \nonumber \\
\frac{d v_x}{dt} & = & - \nu v_x + 4 x \left( 1 - x^2 \right) +
y^2 + f_0 \sin (\omega t) \nonumber \\
\label{os_ODE}
\frac{d y}{dt} & = & 2 v_y \\
\frac{d v_y}{dt} & = & - \nu v_y - 2 y \left( x - p \right) - 4k
y^3 \nonumber
\end{eqnarray}
where $\nu=0.05$, $f_0 = 2.3$, $\omega = 3.5$, $k=0.0075$ and
$p=-1.5$.   This example is interesting for at least two reasons:
\begin{itemize}
\item From the point of view of synchronization, it possess a rich 
structure of invariant manifolds.
\item For the case we will examine,
{\em all} of the important calculations can be performed 
analytically.
\end{itemize}

Originally, Ott and Sommerer examined the two dimensional
invariant manifold defined by $y = v_y = 0$ in the context of
riddled basins.  Their results indicate that, motion on the
manifold is chaotic, however, for 
our parameter values, the manifold is unstable.  Thus, typical
motion for this dynamical systems is in $\R^4$, where there is 
one chaotic attractor and no other attracting sets. 

As usual, denote the drive and response systems by, 
\begin{displaymath}
\frac{d \x}{dt} = \F(\x; t) ,
\end{displaymath}
and
\begin{displaymath}
\frac{d \y}{dt} = \F(\y; t) + \E(\x -\y),
\end{displaymath}
respectively.

In principle the driving trajectory, $\x=[x, v_x, y, v_y] \in
\R^4$, however, we will consider driving trajectories 
restricted to the invariant manifold of the
driving system.  Under these circumstances, $\x = [x, v_x, 0, 0]$
and \DF\ assumes a  block diagonal form.  If we use a block
diagonal coupling matrix, \DE, then the linearized equation of
motion for $\w = \y - \x$ decomposes into motion parallel to, and
perpendicular to, the invariant manifold of the system,
\begin{equation}
\label{two_dim}
\frac{d \w^{(\perp)}}{dt} = \left[ \DF^{(\perp)}(\x) -
\DE^{(\perp)} (\zero) \right] \w^{(\perp)},
\end{equation}
where
\begin{displaymath}
\DF^{(\perp)}(\x) = \left[ \begin{array}{cc} 0  &  1 \\
g^{(\perp)}(x)  & -\nu \end{array} \right] ,
\end{displaymath}
and
\begin{displaymath}
\DE^{(\perp)}(\zero) = \left[
\begin{array}{cc}
\epsilon^{(\perp)}_1 & \epsilon^{(\perp)}_4 \\
\epsilon^{(\perp)}_3 & \epsilon^{(\perp)}_2
\end{array}
\right] ,
\end{displaymath}
and $g^{(\perp)}(x) \equiv -2 (x - p)$. 

An equation similar to Eq.~(\ref{two_dim}) involving
$\w^{(\parallel)}$, $\DF^{(\parallel)}$,  $\DE^{(\parallel)}$, and
$g^{(\parallel)}(x) \equiv 4(1- 3x^2)$ exists for motion parallel
to the manifold.  This decomposition implies, and our numerical
experiments verify, that coupling strengths exist for which the
response trajectory collapses onto the invariant
manifold of the response system 
but the response system is still not synchronized to the
driving system. When this occurs, the response system is linearly
stable to perturbations perpendicular to this manifold but not
to perturbations parallel to this manifold. 

For the remaining discussion we drop the superscripts
$\perp$ and $\parallel$, and leave it to the reader to remember
that each calculation must be performed in {\em both} the
perpendicular and parallel subspaces.  Using Eq.~(\ref{two_dim})
it is easy to show that
\begin{displaymath}
\A = \left[
\begin{array}{cc}
- \epsilon_1 & 1 - \epsilon_4 \\
\left\langle g \right\rangle - \epsilon_3 &  -(\nu + \epsilon_2)
\end{array}
\right],
\end{displaymath}
and
\begin{displaymath}
\B(\x ;t) = \left[
\begin{array}{ccc}
0  & 0 \\
g(x) - \left\langle g \right\rangle & 0
\end{array}
\right].
\end{displaymath}

The eigenvalues of \A\ are
\begin{eqnarray}
\label{os_lambda_pm}
\Lambda_\pm & = & \frac{- (\nu + \epsilon_1 + \epsilon_2)}{2} 
\nonumber \\
 & \pm & \frac{1}{2} \left[ (\nu + \epsilon_2 - \epsilon_1)^2 
+ 4 (1 - \epsilon_4) (\left\langle g \right\rangle - \epsilon_3)
\right]^{1/2},
\end{eqnarray}
and the eigenvectors are
$N_\pm \e_\pm = \left[(1 - \epsilon_4), \: \epsilon_1 +
\Lambda_\pm \right]$, where $N_\pm$ are the following normalizations
$N_\pm = \left[ (1 - \epsilon_4)^2 + |\epsilon_1 + \Lambda_\pm|^2 
\right]^{1/2}$.
The eigenvalues of \A\ can be real or complex, depending on the values
of the $\epsilon$'s.
Using the eigenvectors, $\e_\pm$, and the equation for \B, one
can  obtain the following simple expression (in each subspace)
\begin{eqnarray*}
\lefteqn{\left\langle \| \P^{-1} \left[ \B(\x ;t) \right] \P \|
\right\rangle } \\
 & = & \left\langle | g - \left\langle g \right\rangle |
\right\rangle \left[ \frac{\left( 1 - \epsilon_4
\right)^2}{| \Lambda_{-} - \Lambda_{+} |^2} \right]^{1/2} \left[
\frac{ N_{+}^2 + N_{-}^2}{N_{+} N_{-}} \right].
\end{eqnarray*}

Note that this equation for $\left\langle \| \P^{-1} \left[ \B(\x ;t) 
\right] \P \| \right\rangle$ diverges at the transition between real and
complex $\Lambda_\pm$'s.  From Eq.~(\ref{os_lambda_pm}) it is easy to see 
that $\Re[\Lambda_1]$ increases as the term inside the square root
increases from zero.  
Also, if $\Lambda_\pm$ are complex then $- \Re[\Lambda_1]$ can be
made arbitrarily large by increasing $\epsilon_1$ and/or $\epsilon_2$.
These observations suggest that the best hope for satisfying the
rigorous condition for linear stability of the synchronization
manifold lies with choosing $\epsilon$'s so that
$\Lambda_\pm$ are complex with imaginary parts that are not
too small. 

It is easy to show that if $\Lambda_\pm$ are complex then $N_{+} = N_{-}$
and the rigorous condition for linear stability of synchronization is
\begin{equation}
\label{cond4}
\nu + \epsilon_1 + \epsilon_2 > 4 \left\langle | g - \left\langle
g \right\rangle | \right\rangle C,
\end{equation}
where
\begin{equation}
\label{def_C}
C = \left[ \frac{ - \left(1 - \epsilon_4 \right)^2}{(\nu +
\epsilon_2 -  \epsilon_1)^2 + 4 (1 - \epsilon_4) (\left\langle g
\right\rangle -  \epsilon_3)} \right]^{1/2}.
\end{equation}
(A similar, although more complicated expression for $C$ can be
obtained when $\Lambda_\pm$ are real.)

The major theoretical results of this example are 
Eqs.~(\ref{os_lambda_pm})--(\ref{def_C}), and the conjecture that the
$\epsilon$'s should be chosen so that $\Lambda_\pm$ are complex. 
Together they represent an analytic solution to the rigorous
criteria for synchronization. 

Equation~(\ref{cond4}) is interesting because the right had side
depends on both the coupling strengths and the driving trajectory,
while the left hand side only depends on the coupling strengths.
A useful trick is to set $C = {\rm const.} >0$.   Under these 
circumstances the right hand side of Eq.~(\ref{cond4}) is constant
for a given driving trajectory and one can
increase the value of $\epsilon_1$ and $\epsilon_2$ (maintaining
constant $C$) until the rigorous condition is satisfied.  (As
discussed in Appendix~\ref{geometry2} this approach to designing
couplings that satisfy the rigorous criterion is quite general.)

\begin{figure}
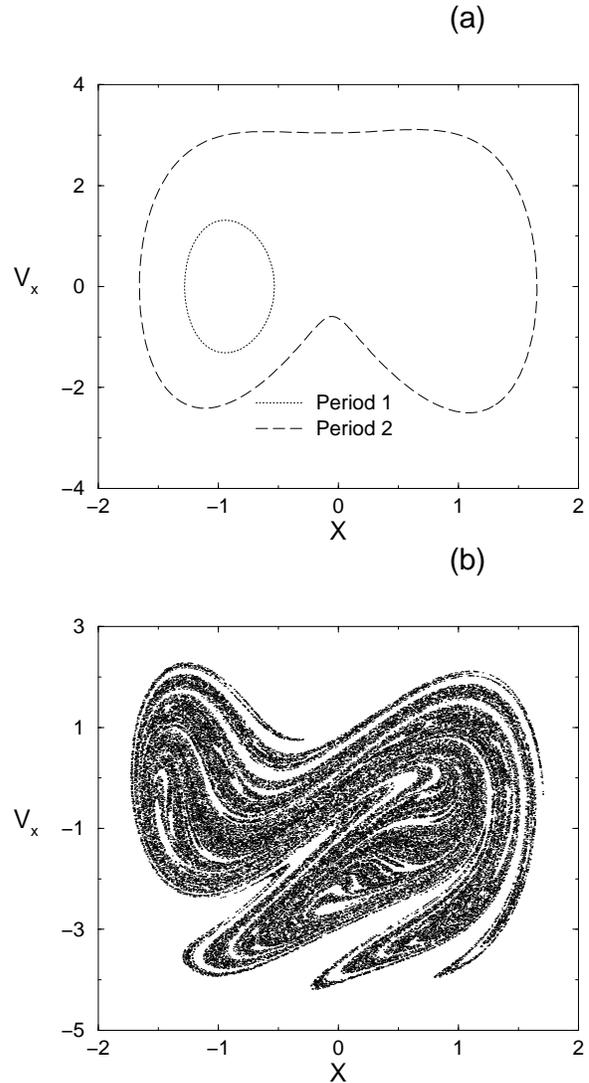

\begin{center}
\leavevmode
\hbox{%
\epsfxsize=3.375in
\epsffile{fig.os.orbits}}
\hbox{%
\epsfxsize=3.375in
\epsffile{fig.os.ss}}
\end{center}
\caption{Measures of the Ott-Sommerer model. (a)  period~1 and
period~2 orbits. (b) the SBR orbit on the surface of section
given by $\omega t = 0$ mod~$2 \pi$.
\label{fig.os.orbits}}
\end{figure}

Because Eqs.~(\ref{os_ODE}) are 
nonautonimous they do not have fixed points.  However, periodic
orbits and the SBR measure still exist.  We examine the SBR and
the Dirac measures associated with a period~1 and a period~2 
orbit (see Figs.~\ref{fig.os.orbits}).  Numerically calculated
values for $\left\langle g \right\rangle$ and $\left\langle |
g - \left\langle g \right\rangle | \right\rangle$ on each of 
these measures are shown in Table~\ref{table.ave}.

We now present examples where the rigorous criterion is tested
for various types of driving.  (Our theoretical and numerical 
results are summarized in Table~\ref{table.os}.)  Our
analysis is simple and is designed to shown whether or not the
rigorous criteria can be satisfied.  Thus the numbers shown in
Table~\ref{table.os} are quite large.  As we show in 
Appendix~\ref{geometry2}, the region of parameter space that
results in couplings that satisfy the rigorous criteria typically
has positive volume.
Thus, with some additional effort we could have found lower values
for the coupling strengths than those listed in Table~\ref{table.os}. 

\subsubsection{Diagonal driving}
This type of driving uses all components of \x\ 
as the driving signal, $\epsilon_3 = \epsilon_4 = 0$, and
$\epsilon_1 = \epsilon_2 \equiv \epsilon$.  For this case the
usable parameter space is the real line \R\ 
and rigorous results have been previously obtained~\cite{fy,hcp}.
For this type of driving Eq.~(\ref{os_lambda_pm}) and 
Table~\ref{table.ave} indicate that $\Lambda_\pm$ are always
complex on each of our measures.  In addition, Eq.~(\ref{def_C})
shows that $C$ is independent of $\epsilon$, and its value is  
fixed by the driving trajectory.  Inserting these results into
Eq.~(\ref{cond4}) leads to the following expression for the
rigorous condition for synchronization
\begin{equation}
\label{test_1}
\epsilon > - \frac{\nu}{2} + 2 \left\langle | g - \left\langle
g \right\rangle |
\right\rangle \left[  \frac{-1}{\nu^2 + 4 \left\langle g
\right\rangle} \right]^{1/2}.
\end{equation}
Since this condition can {\em always} be satisfied linearly
stable synchronization can always be achieved with diagonal driving.

\subsubsection{Driving via position}
This type of driving uses only the position variables, $x$ and $y$.
The simplest example is when $\epsilon_2 = \epsilon_3 = \epsilon_4 
= 0$ and the parameter space is again \R.  In order to demonstrate 
that the rigorous criterion can not be satisfied for this type of 
driving we define new parameters $u \equiv \epsilon_1 + \nu$ and $w
\equiv 1/C$.  In terms of the new parameters Eqs.~(\ref{cond4}) and
(\ref{def_C}) are
\begin{eqnarray*}
u w & > & 4 \left\langle | g - \left\langle g \right\rangle | 
\right\rangle , \\
- 4 \left\langle g \right\rangle & = & (u - 2 \nu)^2 + w^2 .
\end{eqnarray*}
On the measures we are considering $\left\langle g 
\right\rangle < 0$ so these equations define a hyperbola and a circle,
respectively. The circle constrains the values of the new parameters
to a one dimensional curve.  Thus, the dimension of the usable parameter
space is the same for new and old parameters.  ($C$ is imaginary for 
value of $\epsilon_1 \in \R$ corresponding to points off the circle.) 

Synchronization to the driving trajectory
will be linearly stable if any portion of the circle
extends above the hyperbola.  It is straightforward to show that,
on the measures we are examining, the hyperbola has
$w > [- 4 \left\langle g \right\rangle ]^{1/2}$ when $u - 2 \nu = 
[- 4 \left\langle g \right\rangle]^{1/2}$.  Thus the hyperbole and
the circle never intersect and the  rigorous condition for
synchronization can not be satisfied. 
(This result does not mean that this type of driving {\em will not}
produce synchronization.  It only means that our analysis can not
determine a value of $\epsilon_1$ that {\em guarantees}
synchronization.)

Another variation of this type of driving uses the position variables
to drive both the position {\em and} the velocity equation (see 
Ref~\cite{od}).  For this type of driving $\epsilon_2 = 
\epsilon_4 = 0$ and the parameter space, associated with
$\epsilon_1$ and  $\epsilon_3$, is $\R^2$.  Also,
Eq.~(\ref{os_lambda_pm}) and Table~\ref{table.ave} indicate that
$\Lambda_\pm$ are not complex for all values of $\epsilon_1$ and
$\epsilon_3$.  However, if  Eq.~(\ref{def_C}) is satisfied then
$\Lambda_\pm$ are complex.  Rewriting Eqs.~(\ref{cond4}) and
Eq.~(\ref{def_C}) leads to the  following expression for the
rigorous condition for linearly stable synchronization
\begin{eqnarray}
\label{test_2}
\epsilon_1 & > & - \nu + 4 C \left\langle | g - \left\langle g
\right\rangle | \right\rangle, \\
\label{test_3}
\epsilon_3 & = & \frac{1}{4} \left( \nu - \epsilon_1 \right)^2 +
\left[ \left\langle g \right\rangle + \frac{1}{4 C^2} \right], 
\end{eqnarray}
where $C>0$ is arbitrary.  

These equations define a line and a parabola, respectively.
The parabola exist for all possible values of $\epsilon_1$, and all 
driving trajectories, and its minimum value is determined by the
arbitrary constant $C$.  The line also exists for all driving trajectories.
(Of course one should choose $C$ so that this line is in the positive
$\epsilon_1$ half plane.)  These curves are guaranteed to
intersect, and synchronization is guaranteed to be
linearly stable for any point on the parabola whose $\epsilon_1$
value is larger than the one associated with this intersection. 
We remark that $C = 1 / 4 \left\langle | g - \left\langle g \right\rangle |
\right\rangle$ is a choice which greatly simplifies the
guaranteed synchronization condition (this is the form used to 
generate the numerical values shown in Table~\ref{table.os})
\begin{eqnarray*}
\epsilon_1 & > & 1 - \nu \\
\epsilon_3 & = & \frac{1}{4}(\nu - \epsilon_1)^2 + \left[
\left\langle g  \right\rangle + 4 \left\langle | g - \left\langle g \right\rangle |
\right\rangle^2 \right].
\end{eqnarray*}

\subsubsection{Driving via velocity}
This type of driving uses only the velocity
variables, $v_x$ and $v_y$.  As an example, let the velocity
variables drive both the position and the velocity equations.
Thus, $\epsilon_1 = \epsilon_3 = 0$ and the parameter space 
is $\R^2$.   Again we notice that if
Eq.~(\ref{def_C}) is satisfied then 
$\Lambda_\pm$ are complex.  However, 
even if Eq.~(\ref{def_C}) is satisfied, 
values of $\epsilon_2$ and $\epsilon_4$ 
which satisfy the rigorous condition for synchronization do not exit.
To see this define new parameters $u \equiv \nu + \epsilon_2$ and
$w \equiv (\epsilon_4 - 1)/C$.  In terms of the new parameters,
Eqs.~(\ref{cond4}) and (\ref{def_C}) are
\begin{eqnarray*}
u & > & 4 C \left\langle | g - \left\langle g \right\rangle | 
\right\rangle \\
\left( 2 C \left\langle g \right\rangle \right)^2 & = & u^2
+ \left( 2 C \left\langle g \right\rangle - w \right)^2  ,
\end{eqnarray*}
where $C$ is arbitrary.  These equations define a line and
a circle, respectively.  It is straightforward to show that 
the circle does not intersect the line on the measures we have
examined.  Therefore, the rigorous condition for synchronization
can not be satisfied by this type of driving.

\section{Summary}
\label{sum}
In this paper we investigated the linear stability of the invariant 
manifold associated with synchronous behavior between coupled dynamical
systems.  (See Appendix~\ref{geometry1} for a discussion of these
manifolds.)  Although our formalism examined a particular type of coupling
our results are valid for more general types of coupling, and they
can be used to determine the linear stability of invariant manifolds 
within a dynamical system (see Appendix~\ref{other}). 

We have two major results.  The first is a rigorous criterion,
given by Eqs.~(\ref{def_A})--(\ref{cond2}).   If it is
satisfied then linear stability of synchronization to the driving
trajectory is guaranteed.  The condition is based on norms of
deviations from synchronous behavior.  As such it tends to over 
estimate the coupling strength needed to achieve synchronization.
(An alternative formulation of the rigorous criteria can be found
in Appendix~\ref{lorenz}).  The second
result, given by Eq.~(\ref{cond3}), is a ``quick and dirty''
criterion for estimating the coupling strengths need to produce
synchronous behavior.  The approximation
is easy to calculate and can be used to quickly investigate
many coupling schemes and strengths.  A major advantage to our
approach is that both the rigorous and approximate criterion have
a geometric interpretation that can
be used to design couplings schemes and strengths that will result
in synchronization (see Appendix~\ref{geometry2}).  Both criterion
can also be related to fundamental issues of control theory (see
Appendix~\ref{control}). 

Both the
rigorous and approximate condition are dependent on the measure
used for the driving dynamics and can yield different stability
results for different driving trajectories.  Nonetheless, based 
on our numerical results, and previous work by others, it may be
possible to use our criterion to determine couplings that result
in linearly stable synchronization for arbitrary driving 
trajectories~\cite{ho}.

To test these criteria we have performed numerical experiments
on four different dynamical systems, the Rossler, Lorenz (see
also Appendix~\ref{lorenz}), Ott-Sommerer system and a system
used to model chaotic masking in communication (see  
Appendix~\ref{other}).  When taken together,
these examples provide a thorough examination of the stability
criterion we propose.

\begin{figure}
\begin{center}
\leavevmode
\hbox{%
\epsfxsize=3.375in
\epsffile{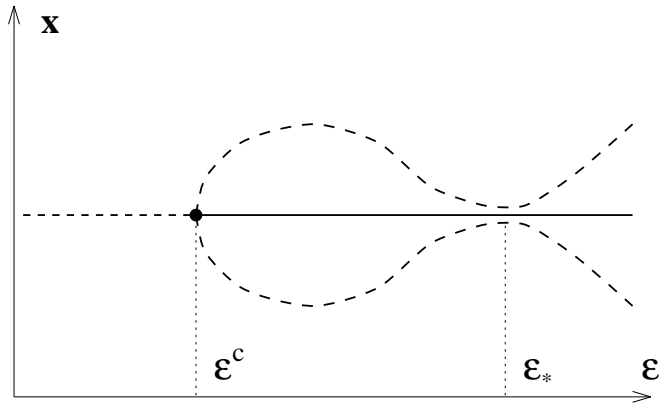}}
\end{center}
\caption{An example of a possible bifurcation diagram for the
stability of synchronization onto an unstable fixed point.  If
the coupling parameter has a value near either $\epsilon^{(c)}$
or $\epsilon_*$ then it is likely that noise will cause the
response system flip beyond the unstable solution and away from
the synchronization manifold.
\label{fig.bif}}
\end{figure}

We close with a discussion of how noise and nonlinear effects
influence the conclusions one can draw from the linear
stability analysis.  In this context consider a chaotic attractor
that contains an unstable fixed point at $\x = \zero$.  Furthermore,
assume that $\x = \zero$ is the driving trajectory, and the response
system synchronizes onto this fixed point as the coupling strength
increases.  The change in stability as
$\epsilon$ increases is a bifurcation which we will model by a
pitchfork, (see Fig~\ref{fig.bif}).  A linear stability analysis does
not take into account the existence of the unstable trajectories
for $\epsilon > \epsilon^{(c)}$.  For arbitrarily small noise
amplitude there exists a range of $\epsilon$ values near
$\epsilon^{(c)}$ where noise will eventually push the
response system above or below  the dashed lines.  When this
occurs the response system can not collapse back to the
synchronized state and is force to seek out an attracting state
away from the synchronization manifold.  Also, a linear stability 
analysis does not take into account the existence of the unstable
trajectories near $\x = \zero$ when $\epsilon \simeq \epsilon_*$.
If $\epsilon \simeq \epsilon_*$ then noise may cause a
loss of synchronization.  Therefore,
although a linear stability analysis
may seem to guarantee stable synchronous motion, noise and nonlinear
effects may prevent long term synchronous behavior. 

\section{Acknowledgments}
The author would like to thank Alistair Mees, David Walker,
Dan Gauthier, Bill Helton, and Lou Pecora for graciously providing
encouragement and advice during the performance of this work. 
The author would like to thank Lev Tsimring and Doug Ridgway for
helpful discussions.  R Brown was supported by the Office of
Naval Research, grant No~N00014-95-1-0864 and the  AirForce
Office of Scientific Research, grant No.~F49620-95-1-0261. N.
Rulkov was supported the Department of Energy, grant 
No.~DE-FG03-95ER14516.

\appendix

\section{Related problems}
\label{other}
With minor modifications the techniques and results discussed
above are valid 
for more general types of coupling.  They can also be used to 
study the stability of invariant manifolds of a dynamical system.
In part, this generality exits because the 
linearized stability equations we examine arise in a variety
of other problems, many of which have recently appeared in the literature.
The key point to realize is that a linearized equation,
similar to Eq.~(\ref{linear}), arises whenever a dynamical system 
has dynamics on a smooth invariant
manifold of its full phase space, and one is considering the linear
stability of this manifold to perturbations which are transverse to
the manifold.

For example, the equations examined by Ott--Sommerer are of the form
\begin{eqnarray*}
\frac{d \x}{dt} & = & \F(\x, \y ;t) \\
\frac{d \y}{dt} & = & \M(\x, \y ;t) \y,
\end{eqnarray*}
where \M\ is an $n \times n$ matrix whose elements are functions of 
$\x \in \R^m$ and $\y \in \R^n$ and time, $t$.  For this system $\y= \zero$
is an invariant manifold.  The linearized equations of motion for $\w =
\y - \zero$ are the same as Eq.~(\ref{two_dim}) with $\DE^{(\perp)} = 
\zero$ (see Eqs.~(7) and (8) in Ref~\cite{os}).  If 
this manifold is unstable then one can use our criterion to determine
a $\DE^{(\perp)}$ that stabilizes this manifold.
By absorbing $\DE^{(\perp)}$ into the matrix $\M(\x, \y ;t)$
one can determine values for the parameters in \M\ that yield a
linearly stable invariant manifold. 

Another type of problem that is covered by our formalism is synchronization
between mutually coupled identical systems.  The equations of motion for
this problem are often of the form
\begin{eqnarray*}
\frac{d \x}{dt} & = & \F(\x) + \E_1(\y-\x) \\
\frac{d \y}{dt} & = & \F(\y) + \E_2(\x-\y) ,
\end{eqnarray*}
where $\E_1$ and $\E_2$ are functions of their argument and
$\E_1(\zero) = \E_2(\zero) = \zero$.  Because of the lack of
constant terms in $\E_1$ and $\E_2$ the synchronization manifold
$\x = \y$ is invariant.  The linear stability of this manifold is
determined by Eq.~(\ref{linear}) where $\DE(\zero) = \DE_1(\zero) +
\DE_2(\zero)$.  Using our criterion
to determine an \E\ that results in linearly stable synchronization is
equivalent to determining forms for $\E_1$ and $\E_2$.  (Here, the nonuniqness
of the decomposition $\DE = \DE_1 + \DE_2$ can probably be reconciled
using details of the specific problem under investigation.) 

Finally, synchronization in drive response systems of the form
\begin{eqnarray*}
\frac{d \x}{dt} & = & \F(\x) \\
\frac{d \y}{dt} & = & \G(\y, \x ; \epsilon) ,
\end{eqnarray*}
where $\F = \G$ when $\x = \y$ and/or $\epsilon = 1$ has been
previously examined~\cite{od,VR93,kp}.  For this system $\x = 
\y$ is an invariant manifold for all values of $\epsilon$.  (We
present an example of this type of system below.)

For each of these problems the important research issues 
centers around the stability of an invariant manifold of the
dynamics.  In particular, synchronization between $N$ coupled
dynamical systems can often  be thought of as motion on a smooth
invariant manifold which lives in the full phase space of a
single large dynamical system consisting of the $N$ smaller
systems.  Given this interpretation, our results indicate
that the stability of synchronization is a special case of
the stability of invariant manifolds of a dynamical system.
(A similar observation can be found in Ref.~\cite{abs}.) 
Therefore, although our results are presented in the
context of synchronization between identical chaotic systems
coupled in a drive/response manner they are applicable to 
areas of research where the stability of invariant
manifolds is the central issue.  Examples of this are 
riddled basins and on-off intermittency, synchronization of 
identical systems with mutual coupling, and generalized 
synchronization~\cite{abs,os,VR93,kp,lg,pst,vaos}.

\subsection{Communication example}
This example is a modification of a synchronization method
which has previously been used to experimentally demonstrate
communication via modulated chaotic signals~\cite{VR93}.  The
modification involves introducing a parameter, $\epsilon$, whose
variation changes the stability properties of the synchronization
manifold.  Introducing this additional parameter, and carefully
selecting its value, can enhance the stability of the synchronization
manifold~\cite{rvrrv94,od}.  Therefore, adding this parameter is very 
useful for applications.

\begin{figure}
\begin{center}
\leavevmode
\hbox{%
\epsfxsize=3.375in
\epsffile{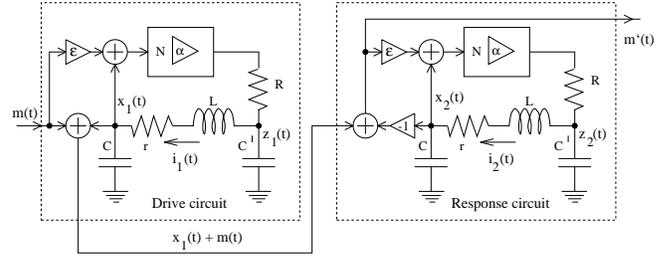}}
\end{center}
\caption{A block diagram of electronic circuits that are modeled
by Eqs.~(\protect \ref{drive_circ}) and (\protect \ref{resp_circ}).
\label{fig.block}}
\end{figure}

A block diagram of the modified drive and response circuits is 
shown in Fig.~\ref{fig.block}.  Each circuit consist of a nonlinear
converter, $N$, which transforms input voltage, $u$, into output, 
$\alpha f(u)$ (see Ref.~\cite{nr} for details).  The parameter
$\alpha$ characterizes the gain of $N$ around $x=0$.  The
nonlinear amplifier has linear feedback  which contains a series
connection to a low-pass filter, $RC'$, and a resonant circuit $LC$.  

Our analysis requires knowledge of the equations of motion
for the systems.   The dynamics of the chaotic driving 
circuit can be described by the following system of three ODE's
\begin{eqnarray}
\frac{d x_{1}}{dt} & = & y_{1} \nonumber \\ 
\label{drive_circ}
\frac{d y_{1}}{dt} & = & -x_{1}- \delta y_{1} + z_{1} \\ 
\frac{d z_{1}}{dt} & = & \gamma[\alpha f(x_{1} + \epsilon m) 
- z_{1}] - \sigma y_{1} , \nonumber
\end{eqnarray}
where the parameters $\alpha$, $\gamma$, $\delta$, $\epsilon$ and 
$\sigma$ are all positive.  (The relations between the parameters 
of the model and the parameters of the circuits can be found in 
Ref.~\cite{nr}). In these equations $m$ denotes the time dependent
message one wants to send, although $x_1 + m$ is the actual 
transmitted signal.

In numerical simulations we use
\begin{eqnarray}
\lefteqn{ f(x) = - {\rm sign}(x) \left( \frac{a}{a - c} \right) } 
\nonumber \\
\label{Nfunc}
 & \times & \left(-a + \left[ \left( \frac{a^2 - c}{a^2} \right) 
(f_1(x)-a)^2 + c \right]^{1/2} \right) 
\end{eqnarray}
where
\begin{equation}
\label{fun.i}
f_1(x) = \left\{
\begin{array}{ll}
|x| & \mbox{if $ |x| \leq a $ } \\
- a \left[ 2 |x| - (b + a) \right] / (b - a) & 
\mbox{if $ a < |x| \leq b $ } \\
-a & \mbox{if $ |x| > b $ }.
\end{array}
\right.
\end{equation}
If we set $a=0.5$, $b=1.8$ and $c=0.03$ then $f(x)$,
as given by Eqs.~(\ref{Nfunc}) and (\ref{fun.i}), fits the 
nonlinearity of the converter, $N$, to within $\approx 2\%$ accuracy. 
The validity of the model (\ref{drive_circ}),(\ref{Nfunc}) has been 
confirmed via synchronization between the real circuit and the 
model~\cite{rnn}.

The response system is driven by the voltage $x_1$ from the drive
system.  Experimentally, it is fed into another circuit which couples 
it to the transmitted signal, $x_1 + m$ (see
Fig.~\ref{fig.block}).   The dynamics of the response circuit is
\begin{eqnarray}
\frac{d x_{2}}{dt} & = & y_{2} \, \nonumber \\
\label{resp_circ}
\frac{d y_{2}}{dt} & = & -x_{2} - \delta y_{2} + z_{2} \\
\frac{d z_{2}}{dt} & = & \gamma [ \alpha f[x_{2} + \epsilon
(x_{1} + m - x_2)] - z_{2}] - \sigma y_{2} \nonumber
\end{eqnarray}

The parameter, $\epsilon$, 
indicates the strength of the coupling between the drive
and response circuit.  Note that if $\epsilon = 0$ then the 
drive and response system are uncoupled.  If $\epsilon = 1$
then a driving signal, $x_1 + m$, is the argument of the 
nonlinearity of the response circuit.  If the drive and 
response circuit synchronize then the response system 
operates as a ``chaos filter'' which can be used to extract
the message from the transmitted signal.  (To see this, 
notice that 
if $\epsilon=1$, and the response system synchronizes to
the drive system, then $x_2 = x_1$ and the message, $m$, can be
recovered by subtracting $x_2$ from transmitted signal.)  More
importantly for applications, if the the two systems synchronize 
for $\epsilon \neq 1$ then this same procedure can still be
used to transmit messages.

An examination of Eqs.~(\ref{drive_circ}) and (\ref{resp_circ})
results in the following equations for \A\ and \B
\begin{displaymath}
\A = \left[
\begin{array}{ccc}
 0 & -1 & 0 \\
-1 & -\delta & 1 \\
 0 & -\sigma & \gamma
\end{array}
\right]
\end{displaymath}
and
\begin{displaymath}
\B(\x_1) = \left[
\begin{array}{ccc}
 0 & 0 & 0 \\
 0 & 0 & 0 \\
g(\x_1) & 0 & 0
\end{array}
\right],
\end{displaymath}
where $g(\x_1) = \alpha \gamma (1 - \epsilon) f'(x_1 + m)$ and 
$f'(x) \equiv d f(x)/d x$.  The characteristic equation for \A\ is
\begin{displaymath}
\Lambda^3 + (\delta + \gamma) \Lambda^2 + (1 + \sigma + \gamma
\delta) \Lambda + \gamma = 0.
\end{displaymath}
This equation is not easily solved for
the eigenvalues.  However, the Routh Hurwitz criteria
indicates that $\Re[\Lambda] < 0$ for all eigenvalues~\cite{rh}.

For this example, \A\ is independent of $\epsilon$
and $\left\langle \| \P^{-1} \left[ \B(\x) \right] \P \| \right\rangle
= \zero$ when
$\epsilon = 1$.  Therefore, if $\epsilon = 1$ then the rigorous 
condition for linear stability of the synchronization
manifold ($\Re[\Lambda_1] < 0$) is satisfied.  If
$\epsilon \neq 1$ then $\left\langle \| \P^{-1} \left[ \B(\x) \right]
\P \| \right\rangle
\neq 0$ and this simple analysis fails.

To understand the importance of $\epsilon \neq 1$ for applications
notice that if $f(x)$ is given by Eqs.~(\ref{Nfunc}) and
(\ref{fun.i}) then $f'(x)$ is not continuous.
However, this function is only an approximation to the real
$f(x)$.  If the real $f(x)$ has a continuous derivative (in fact
we believe $f(x)$ is smooth) then there exists an open neighborhood 
of $\epsilon = 1$ where the rigorous condition for synchronization
(Eq.~(\ref{cond2})) is guaranteed to be satisfied.

The existence of this neighborhood
is important because, although the synchronization manifold
is linearly stable for $\epsilon = 1$ this does not guarantee that 
$\epsilon = 1$ results in the {\em most} stable manifold. 
Because the loss of synchronization is unfavorable for applications
involving communications one wants a synchronization manifold
that is as stable as possible, and which recaptures the response
trajectory as soon as possible in the event that synchronization is
lost.  Therefore, one should to use an $\epsilon$ value that 
yields the {\em most stable} synchronization manifold.

\begin{figure}
\begin{center}
\leavevmode
\hbox{%
\epsfxsize=3.375in
\epsffile{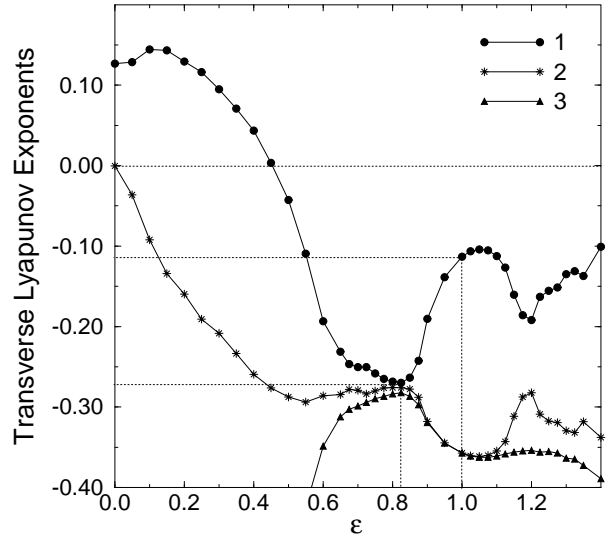}}
\end{center}
\caption{The transverse Lyapunov exponents as a function of
$\epsilon$.
\label{fig.lyap}}
\end{figure}

In Fig.~\ref{fig.lyap} we show the transverse Lyapunov exponents
as a function of $\epsilon$ for our model.  The figure indicates that, 
at least as far as transverse Lyapunov exponents are concerned, the
most stable synchronization manifold occurs when $\epsilon \neq
1$.  The fact that the rigorous condition is guaranteed to be satisfied 
in some neighborhood of $\epsilon = 1$ means that we are free to
search for values of $\epsilon$ that result in the most stable
synchronization manifold.

\section{Geometrical discussion of invariant manifolds}
\label{geometry1}
In this appendix we discusses the various invariant manifolds that arise 
when discussing synchronization.  
The idea of an invariant manifold is fundamental to our approach.
Because we have discussed many different invariant manifolds it is
useful to carefully construct a geometric picture of these manifolds. 
This construction provides a unified interpretation of the issues  
discussed above.  Let $\x \in \R^d$ denote the trajectory of the
driving system and let $\y \in \R^d$ denote the trajectory of the
response system.  Hence, the full phase space that describes the
evolution of the two systems is $\R^{2d}$.  If \z\ denotes a
trajectory in the full phase space, then the synchronization 
manifold for identical systems is defined by ${\cal W}^s \equiv \{ 
\z = (\x, \y) \: | \: \x = \y \}$.

Next, assume that the driving trajectory evolves on an
$m \leq d$~dimensional invariant manifold, ${\cal W}_D$.
This manifold may, or may not, be stable to perturbations.
For example, if the driving trajectory is an unstable fixed 
point then $m=0$, and ${\cal W}_D$ is unstable.  If the driving 
trajectory is an unstable limit cycle then $m=1$, and ${\cal 
W}_D$ is unstable.  However, if the driving trajectory is a 
dense orbit on a chaotic attractor then $m \leq d$ and ${\cal W}_D$
is stable.  (For this case we are not saying that ${\cal W}_D$
is the attractor.  Rather, ${\cal W}_D$ is the manifold that
contains the attractor.)  For all of these cases it is possible
for ${\cal W}_D$ to occupy zero volume in $R^d$.  Finally, if the
system is Hamiltonian and the driving trajectory is chaotic then
$m=d$ and ${\cal W}_D$ occupies positive volume in $R^d$.

In addition, the dynamical system itself may have an invariant manifold,
${\cal W}^0 \subset \R^d$ (which we call an invariant manifold of the 
system.)  An invariant
manifold of the system may contain an infinite number of trajectories.
For example, the Ott--Sommerer system discussed above has an invariant
manifold defined by $y = v_y = 0$, and a chaotic attractor exists in this
manifold~\cite{os}.  (This chaotic set is attracting for points within the
manifold defined by $y = v_y = 0$.)  Therefore, it is possible to 
consider a case where ${\cal W}_D$ is restricted to an invariant
manifold of the system, ${\cal W}_D \subseteq {\cal W}^0$.  However,
in general this is not the case.  An important point to keep in mind is
that the driving trajectory defines ${\cal W}_D$, and each different
driving trajectory (in general) defines a different ${\cal W}_D$.

For identical systems, the response system has 
an identical $m$~dimensional manifold, ${\cal W}_R$.  The manifolds 
${\cal W}_D$ and ${\cal W}_R$ are not the same because ${\cal W}_D$
lives in the phase space of the driving system while ${\cal W}_R$ lives
in the phase space of the response system.

\begin{figure}
\vspace{0.0in}
\begin{center}
\leavevmode
\hbox{%
\epsfxsize=3.375in
\epsffile{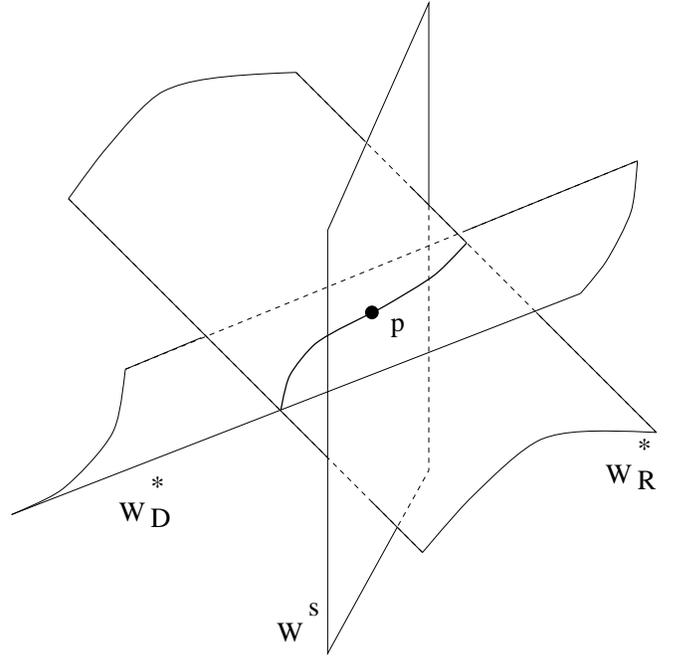}}
\end{center}
\caption{Various manifolds and intersections which arise in
the phase space associated with two identical systems which
are coupled together.
\label{fig.manifolds}}
\end{figure}

In the full phase space of the combined system these manifolds can be
defined by ${\cal W}_D^* \equiv \{ \z = (\x, \y) \: | \: \x \in {\cal W}_D \}$, and
${\cal W}_R^* \equiv \{ \z = (\x, \y) \: | \: \y \in {\cal W}_R \}$. 
(See Fig.~\ref{fig.manifolds} for a schematic picture of these manifolds.)
These manifolds intersect in the full phase space.  The intersection is 
defined by ${\cal W}_D^*  \bigcap {\cal W}_R^* \equiv \{\z = (\x, \y) \: | \: 
\x \in {\cal W}_D \text{ and } \y \in {\cal W}_R \}$, (see the dark 
solid curve in Fig.~\ref{fig.manifolds}).  Since this definition 
does not require $\x = \y$ it is clear that, although ${\cal W}_D^* \bigcap 
{\cal W}_R^*$ contains part of the synchronization manifold, 
${\cal W}_D^* \bigcap {\cal W}_R^* \neq {\cal W}^s$.  Furthermore, 
${\cal W}^s$ contains all possible driving trajectories.  Therefore,
if ${\cal W}_D$ is restricted to an invariant manifold of the system
(${\cal W}_D \subseteq {\cal W}^0$) then  
some trajectories within ${\cal W}^s$ are not in ${\cal W}_D^*$.

Therefore, when we discuss the linear stability of synchronization {\em
for a given driving trajectory} we are actually discussing the stability
of the manifold given by ${\cal W} \equiv {\cal W}_D^* \bigcap {\cal W}_R^* 
\bigcap {\cal W}^s \equiv \{\z =  (\x, \y) \: | \; \x \in {\cal W}_D^* \text{ 
and } \y \in {\cal W}_R^* \text{ and } \x = \y \}$ to perturbations
that are transverse to this manifold.  The manifold associated
with this intersection is labeled $p$ in Fig~\ref{fig.manifolds}.
Although it is shown as a dot one must keep in mind that the dimension
of the manifold, ${\cal W}$, ranges from a low of zero to a high of $d$.
If ${\cal W}_D^* \bigcap {\cal W}_R^* \bigcap {\cal W}^s$ is stable
for all possible driving trajectories we say that ${\cal W}^s$ is stable.

\section{Geometrical interpretation of the stability criteria}
\label{geometry2}
In this appendix we discuss a geometrical interpretation of the
stability criteria discussed in Sections~\ref{theory1} and \ref{theory2}.
Equations~(\ref{cond2}) and (\ref{cond3}) have geometrical interpretations
which can be used to {\em design} couplings that yield stable synchronous
motion.  One can think of the elements of $\DE(\zero)$ as living in a 
$d^2$~dimensional parameter space.  The right hand side of 
Eq.~(\ref{cond2}) defines a function in this parameter space.
Furthermore, this function can be used to define a family of surfaces
in this parameter space.  Explicitly, the family of surfaces $\Sigma_{\bf
B}$ is defined by $\left\langle \| \P^{-1} \left[ \B(\x ; t) \right] \P
\| \right\rangle = C_{\bf B}$, where $C_{\bf B}$ is a constant.  
In a similar fashion the left has side defines a different function and
family of surfaces.  Explicitly, the family of surfaces $\Sigma_\Lambda$
is defined by $- \Re[\Lambda_1] = C_\Lambda$, where $C_\Lambda$
is a constant. 

\begin{figure}
\vspace{0.0in}
\begin{center}
\leavevmode
\hbox{%
\epsfxsize=3.375in
\epsffile{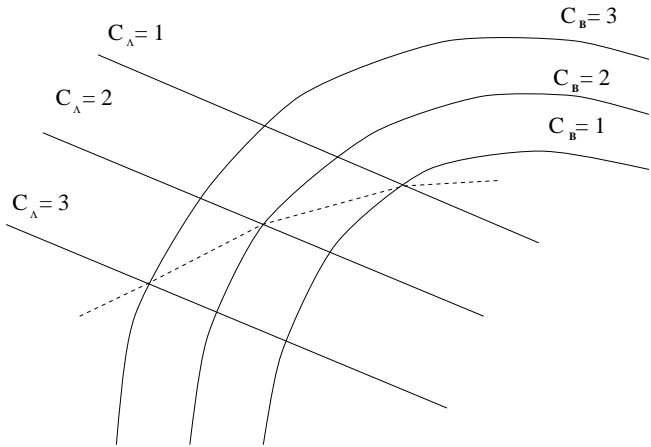}}
\end{center}
\caption{Various surfaces $\Sigma_\Lambda$ and $\Sigma_{\bf B}$. The 
dashed line is the boundary of the region of parameter space that yields
linearly stable synchronization.  Any location in parameter space 
``above'' this line will yield linearly stable synchronization.
\label{fig.surfaces}}
\end{figure}

Now consider a fixed driving trajectory.  For this trajectory, each value 
of $C_{\bf B}$ ($C_\Lambda$) corresponds to a particular surface from the
family $\Sigma_{\bf B}$ ($\Sigma_\Lambda$). 
The boundary of the region of parameter space that yields linearly stable
synchronization is given by the intersections of surfaces from
$\Sigma_\Lambda$ with surfaces from $\Sigma_{\bf B}$ where $C_{\bf B}
= C_\Lambda$.  These intersections form a surface in the parameter space.
Any intersection associated with $C_\Lambda > C_{\bf B}$ will result in
linearly stable synchronization onto the driving trajectory.  Furthermore,
all of these intersections will reside on one side of the surface defined
by the boundary (see Fig.~\ref{fig.surfaces}).  Because $C_\Lambda$ and 
$C_{\bf B}$ are (in general) arbitrary real numbers, and the
reals are dense, we know that the region associated with linearly stable
synchronization will typically occupy positive volume in the phase space.

An immediate consequence of this approach is that if none of the surfaces 
from $\Sigma_\Lambda$ intersect surfaces from $\Sigma_{\bf B}$ then the 
rigorous criterion for synchronization can not be met.  

An explicit examples of this analysis is the Ott--Sommerer example discussed
above.   For this example Eqs.~(\ref{cond4}) and (\ref{def_C}) are equivalent
to the stability criterion of Eq.~(\ref{cond2}). 
Begin with the diagonal driving example and consider a fixed driving
trajectory.  For this example the parameter space is \R\ and $C$  
(from Eq.~(\ref{def_C})) is a constant, independent of $\epsilon$.
Because $C$ is constant, $C_{\bf B}$ can have only one value and there 
is only one surface in the family $\Sigma_{\bf B}$.  Furthermore, the fact
that the right right hand side of Eq.~(\ref{cond4}) is independent of 
$\epsilon$ implies that $\Sigma_{\bf B}$, is the entire 
parameter space.  In contrast $C_\Lambda$ can take on any value.  The
left hand side of Eq.~(\ref{cond4}) indicates that a particular value of
$C_\Lambda$ corresponding to a particular value of $\epsilon$.  Therefore,
$\Sigma_\Lambda$ is a family of points in the parameter space \R.

For these surfaces the boundary of the region of parameter space that yields
linearly stable synchronization (the intersection of $\Sigma_\Lambda$ and
$\Sigma_{\bf B}$ associated with $C_\Lambda = C_{\bf B}$) is a point in \R.
Intersections associated with $C_\Lambda > C_{\bf B}$ are also points.
Finally, the region associated with linearly stable synchronization (the
intersection of all surfaces from $\Sigma_\Lambda$ with surfaces from 
$\Sigma_{\bf B}$ such that $C_\Lambda > C_{\bf B}$) is an interval
which occupies positive volume in \R\ (see Eq.~(\ref{test_1})).

The final example is driving with position, and the position is permitted
to drive both the position and velocity equations.  The parameter space for
this example is $\R^2$.  Now, assume the driving trajectory is fixed.  
The  derivation that lead to Eq.~(\ref{test_3}) shows that, $\Sigma_{\bf B}$
is a family of parabolas in $\R^2$.  Selecting a value for $C_{\bf B}$ is 
equivalent to selecting a value for $C$, which is equivalent to selecting a 
particular parabola from this family.

The left hand side of Eq.~(\ref{cond4}) indicates that $\Sigma_\Lambda$ is a 
family of lines in $R^2$.  Selecting a value for $C_\Lambda$ selects a particular
line from this family.  Thus, if one has chosen a particular value for $C$ then
the boundary of the parameter space region that yields stable synchronization is 
the intersection of a line and a parabola.  For this example this intersection 
always exists.  Intersections of lines associated with $C_\Lambda > C_{\bf B}$ 
(for fixed $C$) sweep out a portion of the parabola.  Therefore, for fixed
$C_{\bf B}$ (equivalently, fixed $C$) the portion of the parabola that is beyond 
the line associated with $C_\Lambda = C_{\bf B}$ corresponds to values of 
the coupling parameters that produce stable synchronous motion.

Now recall that the value of $C$ is arbitrary, and different values of $C$
yield different parabolas.  For each different value of $C$ the 
rigorous criterion chooses a portion of a parabola.  Therefore, the region of
parameter space that satisfies the rigorous criterion is the one swept out by 
portions of the parabolas for all possible values of $C$.  This region
occupies positive volume in the parameter space, $\R^2$.

\section{Relationship to Control Theory}
\label{control}
In this appendix we discuss relationship between our results and those
found in control theory, and alternate formulations of the rigorous
criterion.  The relationship between synchronization and
control has been know for some time~\cite{lg2,mario}.  In these  
papers synchronization is typically discussed as an example of 
feedback control (see the review by Chen and Dong~\cite{cd}). 

In basic control theory one is often interested in the following
equation 
\begin{displaymath}
\frac{d \w}{dt} = \hat{\A}(t) \w + \hat{\B}(t) u(t)
\end{displaymath}
where $u(t)$ is a time dependent scalar input (the drive).  In 
this equation we are invited to think of $\hat{\A}$ as the 
uncontrolled dynamics and $\hat{\B}(t)$ is a vector which couples
the input to the uncontrolled dynamics.  An important question in
control theory is, can one find inputs, $u(t)$, so that the 
$\w(t) = \zero$ in finite time.  The answer to this question is
often determined by constructing a controllability matrix
\begin{equation}
\label{def_G}
G(t, t_0) = \int_{t_0}^t \U(t,s) \hat{\B}(s) \hat{\B}^\dagger(s)
\U^\dagger (t,s) ds ,
\end{equation}
where $\U(t,t_0) = \exp[\int_{t_0}^t \A(s) ds]$.  It is known that
if $\G(t,t_0)$ has full rank ($\det(\G) \neq 0$) then one can find
inputs so that $w(t)=0$~\cite{brogan}.

We are interested in the following linearized equation of motion
\begin{displaymath}
\frac{d \w}{dt} = \left[\DF(t) + \DE(\zero) \right] \w .
\end{displaymath}
For the coupling we have considered the inputs are feedbacks, 
$u(t) = \sum_{\alpha=1}^d \hat{C}_\alpha w_{\alpha}$ where 
$\hat{\C}$ is a constant vector (which typically has only one 
component.)  Therefore, a controllability approach would involve
determining a $\hat{\B}$ so that $\G(t,t_0)$ has full rank and 
then using $DE(\zero)_{\alpha \beta} = \hat{B}_\alpha 
\hat{C}_\beta$ to obtain $\DE(\zero)$.

The difficulty with the controllability approach is that $\U(t,
t_0)$ has a complicated time dependence.  Avoiding this 
complication is one of the motivations for considering the 
equation
\begin{equation}
\label{l1}
\frac{d \w}{dt} = \left[ \A + \B(t) \right] \w ,
\end{equation}
where \A\ and \B\ are defined by Eqs.~(\ref{def_A} and 
(\ref{def_B}).  In this formulation of the problem the coupling
strengths ($\DE(\zero)$) are part of \A, we think of \A\ as the
uncontrolled dynamics, and $\B(t) \w$ are driving inputs.  In 
effect the rigorous criterion of Eq.~(\ref{cond2}) says that if 
the eigenvalues of \A\ are placed sufficiently far into the right 
half plane (as a result of feedback coupling) from the fluctuation  
that result from the chaotic driving will not be sufficient to
destabilize the fixed point at $\w = \zero$.  In this sense our
criterion is similar in spirit to pole placement control theory.

We remark that the controllability matrix can not be immediately 
calculated for Eq.~(\ref{l1}) because $\hat{\B}$ is a function
that maps input space into state space while \B\ is a function
that maps state space into itself.  In Eq.~(\ref{l1}) the 
function that plays the role of $\hat{\B}$ is part of the matrix
\A.

We believe that a different interpretation of synchronization
(still from control theory) could be quite promising.  Walker and 
Mees interpret synchronization as an example state estimation and
the observer problem~\cite{wm}.  This view
of synchronization says that synchronization permits one to estimate
the complete state of a system when the full state vector is not
observable.  In practice one could imagine coupling the output from
a physical system to a model of the system using coupling that is
guaranteed to result in stable synchronization.  (Here we rely 
on the fact that synchronization is robust to modeling errors and
noise~\cite{rnn}).  The full state of the physical is then the 
same as the state of the model.
For linear system the observer problem and the control problem can
can be shown to be duals of each other~\cite{sontag,brogan}.  Thus,
many of the results proved for observability of a dynamical system 
are corollaries of theorems proved for control.  In this paper a 
rigorous and easy to calculate criterion for synchronization is given.

\section{Another rigorous criterion}
\label{lorenz}

In this appendix we describe a derivation that leads to a
rigorous criterion similar to the one derived in
Section~\ref{theory1}.  If one does not transform to coordinates
given by the eigenvectors of \A\ then Eq.~(\ref{int_eq}) becomes
\begin{displaymath}
\w(t) = \U(t, t_0) \w(t_0) + \int_{t_0}^t \U(t,s) \B(s) \w(s) ds,
\end{displaymath} 
where $\U(t, t_0) = \exp[\A (t - t_0)]$.  This equation leads to
Eq.~(\ref{cond1}) as a sufficient condition for linear stability
of synchronization with \K\ replaced by \B.  It is straightforward
to shown if $\U(t, t_0) = \exp[\A (t - t_0)]$ then a rigorous 
sufficient condition for linear stability of synchronization is
\begin{equation}
\label{new_cond}
-\Re[\Lambda_1] > C \left\langle \| \B \| \right\rangle.
\end{equation}
where
\begin{equation}
\label{new_C}
C =  \left[ \sum_{\alpha \beta} | P_{\alpha 1} P_{1 
\beta}^{-1} |^2 \right]^{1/2} .
\end{equation}
For this criterion the terms involving the coupling
strength and the time average have been decoupled.  This could
greatly simplify the effort required for some calculations.  

As an example of this consider the Lorenz system. The eigenvectors
of \A\ are
\begin{eqnarray*}
\e_z & = & \hat{z} \\ 
\e_\pm & = & \frac{1}{N_\pm} \left[(\sigma - \epsilon_4), \:
(\sigma + \epsilon_1 + \Lambda_\pm), \: 0 \right],
\end{eqnarray*}
where $N_\pm$ are the following normalizations
\begin{displaymath}
N_\pm = \left[ (\sigma - \epsilon_4)^2 + |\sigma + \epsilon_1 
+ \Lambda_\pm|^2 \right]^{1/2}.
\end{displaymath}
Also, $\B(\x )$ is given by
\begin{displaymath}
\B(\x) = \left[
\begin{array}{ccc}
 0 & 0 & 0 \\
- z + \left\langle z \right\rangle  & 0 &
- x + \left\langle x \right\rangle \\
  y - \left\langle y \right\rangle  & x - \left\langle x 
 \right\rangle & 0 
\end{array}
\right].
\end{displaymath}

Clearly, calculating $\| \P^{-1} \B(\x) \P \|$ is a complicated
procedure.  However, the rigorous condition of Eq.~(\ref{new_cond})
can be easily satisfied.  It is possible to show that if $\Lambda_\pm$
are complex then
\begin{displaymath}
C = \left[ \frac{(r - \left\langle z \right\rangle - 
\epsilon_3 + 1 )^2}{-[\sigma + \epsilon_1 - 1 - \epsilon_2]^2 -
4 (r - \left\langle z \right\rangle - \epsilon_3) (\sigma
- \epsilon_4)} \right]^{1/2}. 
\end{displaymath}
Now consider off diagonal $y$-driving ($\epsilon_1 = \epsilon_3
= 0$) and fixed $C$.  To satisfy the rigorous condition of 
Eq.~(\ref{new_cond}) while keeping $C$ fixed one needs
\begin{eqnarray*}
\epsilon_4 & = & \frac{(\epsilon_2 + 1 - \sigma)^2}{4(r - 
\left\langle z \right\rangle)} + \left[ \sigma + \frac{(r - 
\left\langle z \right\rangle -1 )^2}{4 C^2 (r - \left\langle z 
\right\rangle)} \right] \\
\epsilon_2 & > & - (\sigma + 1) + 4C\left\langle \| \B \| 
\right\rangle ,
\end{eqnarray*} 
where $C$ is arbitrary.  The second equation is a parabola which
exists for all values of $\epsilon_2$, and all driving trajectories.
The first equation is a line which is guaranteed to intersect the
parabola.  Therefore the rigorous condition of Eq.~(\ref{new_cond})
can be satisfied for this type of driving.

\clearpage

\widetext

\begin{table}
\begin{tabular}{cccccc}
\multicolumn{6}{c}{Rossler System} \\
Measure Type & Drive Type &
\multicolumn{2}{c}{Approximate Test} & 
\multicolumn{2}{c}{Numerical Test} \\
          &   & $\epsilon^{(c)}_{\min}$ & $\epsilon^{(c)}_{\max}$
              & $\epsilon^{(c)}_{\min}$ & $\epsilon^{(c)}_{\max}$
                \\ 
\tableline
              & $x$ & 0.1975 & 4.998    & 0.1975 & 4.998 \\
Fixed Point   & $y$ & 0.1976 & $\infty$ & 0.1976 & $\infty$ \\
              & $z$ & 9.0    & 225      & 9.0    & 225  \\
\multicolumn{6}{c}{  } \\
              & $x$ & 0.04   & 4.8      & 0.54   & 3.7 \\
Period~1      & $y$ & 0.04   & $\infty$ & 0.36   & $\infty$  \\ 
              & $z$ &   F    &    F     & \multicolumn{2}{c}{None} \\
\multicolumn{6}{c}{ } \\
              & $x$ & 0.07   & 4.9      & 0.32   & 4.3 \\
Period~2      & $y$ & 0.07   & $\infty$ & 0.29   & $\infty$ \\
              & $z$ &   F    &    F     & \multicolumn{2}{c}{None} \\
\multicolumn{6}{c}{ } \\
              & $x$ & 0.11   & 4.9      & 0.20   & 4.5 \\
SBR           & $y$ & 0.11   & $\infty$ & 0.18   & $\infty$ \\
              & $z$ &   F    &    F     & \multicolumn{2}{c}{None} \\
\end{tabular}
\caption{Results of numerical tests on the Rossler system.  In this
table F implies that this type of driving fails the test, while none
implies that synchronization did not occur.
\label{table.ross}}
\end{table}

\begin{table}
\begin{tabular}{cccccccc}
\multicolumn{2}{c}{              } &
\multicolumn{3}{c}{Rossler System} &
\multicolumn{3}{c}{Lorenz System} \\
\multicolumn{2}{c}{Measure Type} & $\left\langle x \right\rangle$ &
  $\left\langle y \right\rangle$ & $\left\langle z \right\rangle$ &
  $\left\langle x \right\rangle$ & $\left\langle y \right\rangle$ &
  $\left\langle z \right\rangle$ \\
\tableline
\multicolumn{2}{c}{Period 1}  & 0.2770 & -1.385  & 1.385  & 0 & 0 & 54.81 \\ 
\multicolumn{2}{c}{Period 2}  & 0.2246 & -1.123  & 1.123  & 0 & 0 & 54.94 \\
\multicolumn{2}{c}{SBR}       & 0.1649 & -0.8245 & 0.8245 & 0 & 0 & 54.82
\end{tabular}
\caption{Numerical estimates for averages along driving trajectories.
For the Rossler system $\left\langle x \right\rangle = -a \left\langle
y \right\rangle$ and $\left\langle z \right\rangle = - \left\langle
y \right\rangle$.
\label{table.averages}}
\end{table}

\begin{table}
\begin{tabular}{cccccc}
\multicolumn{6}{c}{Lorenz System} \\
Measure Type & Drive Type &
\multicolumn{2}{c}{Approximate Test} & 
\multicolumn{2}{c}{Numerical Test} \\
          &   & $\epsilon^{(c)}_{\min}$ & $\epsilon^{(c)}_{\max}$
              & $\epsilon^{(c)}_{\min}$ & $\epsilon^{(c)}_{\max}$
                \\ 
\tableline
              & $x$ & 590    & $\infty$ & 590    & $\infty$ \\
Fixed Point   & $y$ & 59     & $\infty$ & 59     & $\infty$ \\
              & $z$ &   F    &    F     &  \multicolumn{2}{c}{None} \\
\multicolumn{6}{c}{  } \\
              & $x$ & 42     & $\infty$ & 17     & $\infty$ \\
Period~1      & $y$ & 4.2    & $\infty$ & 4.1    & $\infty$ \\ 
              & $z$ &   F    &    F     & \multicolumn{2}{c}{None} \\
\multicolumn{6}{c}{ } \\
              & $x$ & 42     & $\infty$ & 18     & $\infty$ \\
Period~2      & $y$ & 4.2    & $\infty$ & 4.0    & $\infty$ \\
              & $z$ &   F    &    F     & \multicolumn{2}{c}{None} \\
\multicolumn{6}{c}{ } \\
              & $x$ & 42     & $\infty$ & 19     & $\infty$ \\
SBR           & $y$ & 4.2    & $\infty$ & 4.0    & $\infty$ \\
              & $z$ &   F    &    F     & \multicolumn{2}{c}{None} \\
\end{tabular}
\caption{Results of numerical tests on the Lorenz system.  In this
table F implies that this type of driving fails the test, while none
implies that synchronization did not occur.
\label{table.lor}}
\end{table}

\begin{table}
\begin{tabular}{cccccc}
\multicolumn{6}{c}{Ott-Sommerer System} \\
\multicolumn{2}{c}{Measure Type}  & $\left\langle g^{(\perp)}
\right\rangle$ & $\left\langle g^{(\parallel)} \right\rangle$ &
$\left\langle | g^{(\perp)} - \left\langle g^{(\perp)} \right\rangle 
| \right\rangle$ & $\left\langle | g^{(\parallel)} - \left\langle
g^{(\parallel)} \right\rangle | \right\rangle$ \\
\tableline
\multicolumn{2}{c}{Period 1}  & -1.223 & -6.307 & 0.4769 &
5.142 \\
\multicolumn{2}{c}{Period 2}  & -3 & -7.767 & 1.678  & 10.30 \\
\multicolumn{2}{c}{SBR}  & -3 & -7.038 & 1.714  & 7.856 \\
\end{tabular}
\caption{\label{table.ave}}
\end{table}

\clearpage

\begin{table}
\begin{tabular}{cccccccc}
\multicolumn{8}{c}{Ott-Sommerer System} \\
Drive Measure & Drive Type &
\multicolumn{2}{c}{Rigorous Tests} &
\multicolumn{2}{c}{Approximate Test} & 
\multicolumn{2}{c}{Numerical Test} \\
 & & $\perp$ & $\parallel$ & $\perp$ & $\parallel$ & $\perp$ &
$\parallel$ \\
\tableline
Period 1      & Diagonal All          & 0.812 & 4.04 & 0 & 0 & 0.01 &
0.28 \\
              & Diagonal only $x$     &   F   &   F  & 0 & 0 & 0.01 &
0.62 \\
	      & Diagonal only $v$     &   F   &   F  & 0 & 0 & 0.01 &
0.69 \\
              & Off Diagonal $x$      & 0.95  & 0.95 & 0 & 0 & 0.002&
0.15 \\
              &                       &   0   & 99.6 & 0 & 0 & 0.04 &
0.55 \\
\multicolumn{8}{c}{ } \\
Period 2      & Diagonal All          & 1.89  & 7.34 & 0 & 0 & 0.13 &
0.99 \\
              & Diagonal only $x$     &   F   &   F  & 0 & 0 & 0.25 &
2.6  \\
              & Diagonal only $v$     &   F   &   F  & 0 & 0 & 0.24 &
2.8  \\
              & Off Diagonal $x$      & 0.95  & 0.95 & 0 & 0 & 0.05 &
0.5  \\
              &                       & 8.47  & 416  & 0 & 0 & 0.9  &
1.75 \\
\multicolumn{8}{c}{ } \\
SBR           & Diagonal All          & 1.92  & 5.87 & 0 & 0 & 0.05 &
0.52 \\
              & Diagonal only $x$     &   F   &   F  & 0 & 0 & 0.09 &
1.9  \\
              & Diagonal only $v$     &   F   &   F  & 0 & 0 & 0.10 &
1.5  \\
              & Off Diagonal $x$      & 0.95  & 0.95 & 0 & 0 & 0.05 &
0.5  \\
              &                       & 8.95  & 193  & 0 & 0 & 0.7  &
1.75 \\
\end{tabular}
\caption{Results of numerical tests on the Ott-Sommerer system. 
In this table F implies that this type of driving fails the test.
For off diagonal driving the first number listed is either 
$\epsilon_1$ or $\epsilon_2$ while the number listed below it is
either $\epsilon_3$ or $\epsilon_4$.
\label{table.os}}
\end{table}

\end{document}